\documentclass[twocolumn]{autart}
\usepackage{amsmath}
\usepackage{amssymb}
\usepackage{mathtools}
\usepackage{graphicx}
\usepackage{soul,color} 
\usepackage{citesort} 
\usepackage{tikz} 
\usepackage{enumerate}
\usepackage[algo2e,norelsize,linesnumbered,ruled]{algorithm2e}
\usepackage[linesnumbered,ruled]{algorithm2e}
\usepackage{array}\newcolumntype{M}[1]{>{\centering\arraybackslash}m{#1}}
\usepackage{flushend} 
\usepackage{balance} 
\usepackage{changes}
\usepackage{marginnote}

\begin{document}

\begin{frontmatter}
	
	\title{Set operations and order reductions for constrained zonotopes \thanksref{footnoteinfo}}
	\thanks[footnoteinfo]{This material is based upon work supported by the National Science Foundation under Grant No. 1849500. This paper was not presented at any IFAC meeting. Corresponding author J. P. Koeln Tel. +1 972-883-4649. Fax +1 972-883-4659.}
	
	\author{Vignesh Raghuraman}\ead{vignesh.raghuraman@utdallas.edu},
	\author{Justin P. Koeln}\ead{justin.koeln@utdallas.edu}
	
	\address{University of Texas at Dallas, Richardson, TX, 75080, United States}
	
	\begin{abstract}        
		This paper presents methods for using zonotopes and constrained zonotopes to improve the practicality of a wide variety of set-based operations commonly used in control theory. The proposed methods extend the use of constrained zonotopes to represent sets resulting from operations including halfspace intersections, convex hulls, robust positively invariant sets, and Pontryagin differences. Order reduction techniques are also presented that provide lower-complexity inner-approximations of zonotopes and constrained zonotopes. Numerical examples are used to demonstrate the efficacy and computational advantages of using zonotope-based set representations for dynamic system analysis and control.
	\end{abstract}       
	
	
\end{frontmatter}

\section{Introduction}

Sets are widely used in a variety of control theory and applications including reachability analysis for system verification \cite{Asarin2006,Girard2005,Girard2006,Kurzhanskiy2007}, robust Model Predictive Control (MPC) \cite{Mayne2005,Langson2004,Bravo2006}, and state estimation \cite{Chisci1996,Alamo2005,Le2013}. However, the sets used in control theory are not always practical to compute in application. For example, the minimal Robust Positively Invariant (mRPI) set \cite{Rakovic2005} is widely used in robust MPC \cite{Mayne2014,Richards2006,Limon2010}.  However, in general, mRPI sets are not finitely represented and must be approximated. Furthermore, existing techniques for determining finite approximations of the mRPI set do not scale well with the dimension of the state space. Such scalability issues are found in many set computations \cite{Tiwary2008}, motivating the need for alternative set representations and efficient approximation algorithms. 

When computing a set, there is often a trade-off between accuracy, complexity, and computation time. The desired balance of these three aspects varies depending on if the set computations are performed off-line prior to controller execution or on-line in real-time. Certain applications permit iterative set computation methods while others require one-step methods that allow set computations to be embedded within an existing optimization problem \cite{Trodden2016_OneStep}.

Additionally, this trade-off is highly dependent on the specific representation of the set. Widely used set representations include the halfspace representation (H-Rep) based on the intersection of a finite number of halfspace inequalities and the vertex representation (V-Rep) based on the convex hull of a finite number of vertices. As an alternative, zonotopes (G-Rep) \cite{McMullen1971} and, more recently, constrained zonotopes (CG-Rep) \cite{Scott2016} have enabled significant reductions in the cost and complexity associated with commonly used set computations in dynamic systems and control.

A \emph{zonotope} is the Minkowski sum of a finite set of line segments or, equivalently, the image of a hypercube under an affine transformation \cite{Fukuda2004,Maler2008}. Due to their computational efficiency, zonotopes have been widely used in reach set calculations for hybrid system verification, estimation, and MPC \cite{Maler2008,Althoff2010,Scott2016,Bravo2006}. As with the iterative algorithm in \cite{Scibilia2011}, computing these reach sets utilizes linear transformation and Minkowski sum operations. Zonotopes are closed under these operations (i.e. the Minkowski sum of two zonotopes is a zonotope) and the number of generators grows linearly with the number of Minkowski sum operations, compared to the potential exponential growth of the number of halfspaces in H-Rep. Unfortunately, zonotopes in general are not closed under intersection and the conversion from G-Rep to H-Rep for intersection operations is inefficient. 

\emph{Constrained zonotopes} were developed in \cite{Scott2016} to overcome the limitations caused by the inherent symmetry of zonotopes. Constrained zonotopes are closed under linear transformation, Minkowski sum, and generalized intersection and can be used to represent any convex polytope. Constrained zonotopes provide the computational advantages of zonotopes while enabling exact computations of a much wider class of sets. In \cite{Koeln2019ACC}, reach set computations using constrained zonotopes were shown to be several orders-of-magnitude faster than the same set computations using H-Rep, enabling the on-line computation of these reach sets for use in a hierarchical MPC formulation.

While zonotopes and constrained zonotopes provide a significant computational advantage, various set operations can increase the complexity of the resultant sets beyond a desired upper limit. Thus, there is a need for techniques that provide reduced-complexity approximations of the desired set. Currently there exist reduced-order outer-approximation techniques for zonotopes \cite{Kopetzki2017a,Yang2018} and constrained zonotopes \cite{Scott2016}.  Outer-approximations are widely used in the field of reachability analysis for system verification to determine if a system will always operate in a desired region of the state space \cite{Girard2005,Girard2006}.  

However, in many applications there is a need for computing reduced-order inner-approximations. In general computing inner-approximations of sets is considered a more difficult problem \cite{Kurzhanski2000}. Inner-approximations are particularly important when computing backward reachable sets that define a set of initial states for which a system will enter a specified target region after some allotted time \cite{Xue2017}. While there are existing techniques for zonotopes \cite{Girard2006,Han2016a}, inner-approximation techniques for constrained zonotopes are lacking.

The goal of this paper is to further increase the practicality of applying set-based control techniques through the use of zonotopes and constrained zonotopes. Specifically, this paper provides improved methods for i) representing set intersections with halfspaces, ii) removing redundancy from set representations, and iii) computing reduced-order inner-approximations, convex hulls, RPI sets, and Pontryagin differences. Approaches for both zonotopes and constrained zonotopes are provided along with numerical examples that demonstrate the features and applicability of each approach.\footnote{The source code for all of the constrained zonotope operations and numerical examples is provided at https://github.com/ESCL-at-UTD/ConZono.}

The remainder of the paper is organized as follows. Section~\ref{Notation} provides some initial notation and preliminary background on set operations, zonotopes, and constrained zonotopes. Methods for checking and computing halfspace intersections for zonotopes and constrained zonotopes are presented in Section~\ref{halfspaceIntersections}. Section~\ref{Sec_Redundancy} addresses the issue of redundancy in set representations along with methods for redundancy removal. Techniques for computing reduced-complexity inner-approximations of zonotopes and constrained zonotopes are provided in Section~\ref{Sec_InnerApprox}. Zonotope and constrained-zonotope based methods for computing the convex hull of two sets, the outer-approximation of the mRPI set, and the Pontryagin difference of two sets are presented in Sections \ref{Sec_ConvexHull}, \ref{Sec_RPI}, and \ref{Sec_Pontryagin}, respectively. Section~\ref{Sec_Hier} provides a practical application of these techniques for computing and approximating a backward reachable set within the context of hierarchical control. Finally, Section~\ref{Conclusions} summarizes the conclusions of the paper.
\section{Notation and Preliminaries} \label{Notation}

For sets $ Z, W \subset \mathbb{R}^n $, $ Y \subset \mathbb{R}^m $, and matrix $ \mathbf{R} \in \mathbb{R}^{m \times n} $, the linear transformation of $ Z $ under $ \mathbf{R} $ is $ \mathbf{R} Z = \left\{\mathbf{R} \mathbf{z} \mid \mathbf{z} \in Z \right\} $, the Minkowski sum of $ Z $ and $ W $ is $ Z \oplus W = \left\{\mathbf{z}+\mathbf{w} \mid \mathbf{z} \in Z, \mathbf{w} \in W \right\} $, and the generalized intersection of $ Z $ and $ Y $ under $ \mathbf{R} $ is $ Z \cap_{\mathbf{R}} Y = \left\{ \mathbf{z} \in Z \mid \mathbf{R}\mathbf{z} \in Y \right\} $. The standard intersection, corresponding to the identity matrix $ \mathbf{R} = \mathbf{I}_n $, is simply denoted as $ Z \cap Y$. 

The convex polytope $ H \subset \mathbb{R}^n $ in H-Rep is defined as $ H = \{\mathbf{x} \in \mathbb{R}^n \mid \mathbf{H} \mathbf{x} \leq \mathbf{f} \} $ where $ \mathbf{H} \in \mathbb{R}^{n_h \times n} $, $ \mathbf{f} \in \mathbb{R}^{n_h} $, and $ n_h $ is the number of halfspaces. A centrally symmetric set $ Z \subset \mathbb{R}^n $ can be represented as a zonotope in G-Rep where $ Z = \left\{ \mathbf{G} \boldsymbol{\xi} + \mathbf{c} \mid \lVert \boldsymbol{\xi} \rVert_\infty \leq 1 \right\} $. The vector $ \mathbf{c} \in \mathbb{R}^n $ is the center and the $ n_g $ generators, denoted $ \mathbf{g}_i $, form the columns of the generator matrix $ \mathbf{G} \in \mathbb{R}^{n \times n_g } $. Similarly, a constrained zonotope $ Z_c \subset \mathbb{R}^n $ is defined in CG-Rep as $ Z = \left\{ \mathbf{G}\boldsymbol{\xi} + \mathbf{c} \mid \lVert \boldsymbol{\xi} \rVert_\infty \leq 1, \mathbf{A} \boldsymbol{\xi} = \mathbf{b} \right\} $. With $ \mathbf{A} \in \mathbb{R}^{n_c \times n_g} $ and $ \mathbf{b} \in \mathbb{R}^{n_c} $, constrained zonotopes include $ n_c $ equality constraints that break the symmetry of zonotopes and allow any convex polytope to be written in CG-Rep. The complexity of a zonotope is captured by its order, $ o = \frac{n_g}{n} $ while the complexity of a constrained zonotope is captured by the degrees-of-freedom order, $o_d = \frac{n_g-n_c}{n} $. Zonotopes and constrained zonotopes are denoted as $ Z = \left\{\mathbf{G},\mathbf{c}\right\} $  and $ Z_c = \left\{\mathbf{G},\mathbf{c},\mathbf{A},\mathbf{b}\right\} $, respectively. 

As shown in \cite{Scott2016}, constrained zonotopes are closed under linear transformation, Minkowski sum, and generalized intersection where

\begin{equation} \label{affineMap}
\mathbf{R} Z = \left\{\mathbf{R} \mathbf{G}_z, \mathbf{R} \mathbf{c}_z, \mathbf{A}_z, \mathbf{b}_z\right\},
\end{equation}
\begin{equation} \label{MinkowskiSum}
Z \oplus W = \left\{\left[\mathbf{G}_z \; \mathbf{G}_w\right], \mathbf{c}_z+\mathbf{c}_w, \begin{bmatrix} \mathbf{A}_z & \mathbf{0} \\ \mathbf{0} & \mathbf{A}_w \end{bmatrix}, \begin{bmatrix} \mathbf{b}_z \\ \textbf{b}_w \end{bmatrix} \right\},
\end{equation}
\begin{equation} \label{generalized_Intersection}
Z \cap_{\mathbf{R}} Y = \left\{\left[\mathbf{G}_z \; \mathbf{0}\right], \mathbf{c}_z,  \begin{bmatrix} \mathbf{A}_z & \mathbf{0} \\ \mathbf{0} & \mathbf{A}_y \\ {\scriptstyle \mathbf{R} \mathbf{G}_z} & {\scriptstyle -\mathbf{G}_y} \end{bmatrix}, \begin{bmatrix} \mathbf{b}_z \\ \mathbf{b}_y \\ {\scriptstyle \mathbf{c}_y - \mathbf{R} \mathbf{c}_z }\end{bmatrix} \right\}.
\end{equation}
Additional notation is defined as follows. The set of non-negative real numbers is denoted as $ \mathbb{R}_+ $. The matrix $ \mathbf{T} \in \mathbb{R}^{n \times m} $ with values $ t_{i,j} $ in the $i^{th}$ row and $ j^{th} $ column is denoted as $ \mathbf{T} = [t_{i,j}] $. A $ n \times m $ matrix of zeros is denoted as $ \mathbf{0}_{n \times m} $ or simply $ \mathbf{0} $ if the dimension can be readily determined from context. Similarly, a vector of ones is denoted as $ \mathbf{1} $. For a matrix $ \mathbf{A} $, the null space is denoted $ \mathcal{N}(\mathbf{A}) $ and the pseudoinverse is denoted $ \mathbf{A}^\dagger $. Parallel vectors $ \mathbf{v}_1 $ and $ \mathbf{v}_2 $ are denoted as $ \mathbf{v}_1 \parallel  \mathbf{v}_2 $. The unit hypercube in $ \mathbb{R}^n $ is defined as $ B_\infty = \left\{ \boldsymbol{\xi} \mid \| \boldsymbol{\xi} \|_\infty \leq 1 \right\}$ while $ B_\infty(\mathbf{A},\mathbf{b}) = \left\{ \boldsymbol{\xi} \in  B_\infty \mid \mathbf{A} \boldsymbol{\xi} = \mathbf{b} \right\} $. With the volume of a set $ X $ denoted as $ V(X) $, the volume ratio for sets $ X, Y \in \mathbb{R}^n $ is defined as $ V_r = \left( \frac{V(X)}{V(Y)} \right)^{1/n} $. All numerical examples were generated using MATLAB on a desktop computer with a 3.6 GHz i7 processor and 16 GB of RAM. 
All optimization problems were formulated and solved with YALMIP \cite{Lofberg2004} and Gurobi \cite{Gurobi2019}.

\section{Halfspace Intersections} \label{halfspaceIntersections}

This section presents methods for determining if a zonotope or constrained zonotope intersects a given halfspace along with the exact representation of this intersection in CG-Rep. The need for computing this intersection arises in reachability analysis \cite{Althoff2012} and in MPC when determining the set of feasible initial conditions \cite{Scibilia2011}. The use of CG-Rep enables exact representations unlike existing techniques that rely on zonotopic approximations of the intersection \cite{Girard2008}.

\subsection{Zonotope-Halfspace Intersection}
For a zonotope in $ \mathbb{R}^n $ with $ n_g $ generators, the intersection between a zonotope and a hyperplane can be tested algebraically with complexity $O(nn_g)$. 
\begin{lem} (Section 5.1 of \emph{\cite{Girard2005}}) \label{Zono_halfspace_int_check}
	The zonotope $ Z = \{\mathbf{G},\mathbf{c}\} \subset \mathbb{R}^n $ intersects the hyperplane $ H = \{\mathbf{x} \in \mathbb{R}^n \mid \mathbf{h}^T \mathbf{x} = f \} $ if and only if 
	\begin{equation} \label{hyperplane_int}
	| f - \mathbf{h}^T \mathbf{c} | \leq \sum_{i = 1}^{n_g} | \mathbf{h}^T \mathbf{g}_i |.
	\end{equation} 
\end{lem}
If a zonotope intersects a hyperplane, the intersection between the zonotope and the corresponding halfspace can be represented in CG-Rep by the addition of exactly one generator and one equality constraint. 

\begin{figure}
	\begin{center}
	\includegraphics[width=8.6cm]{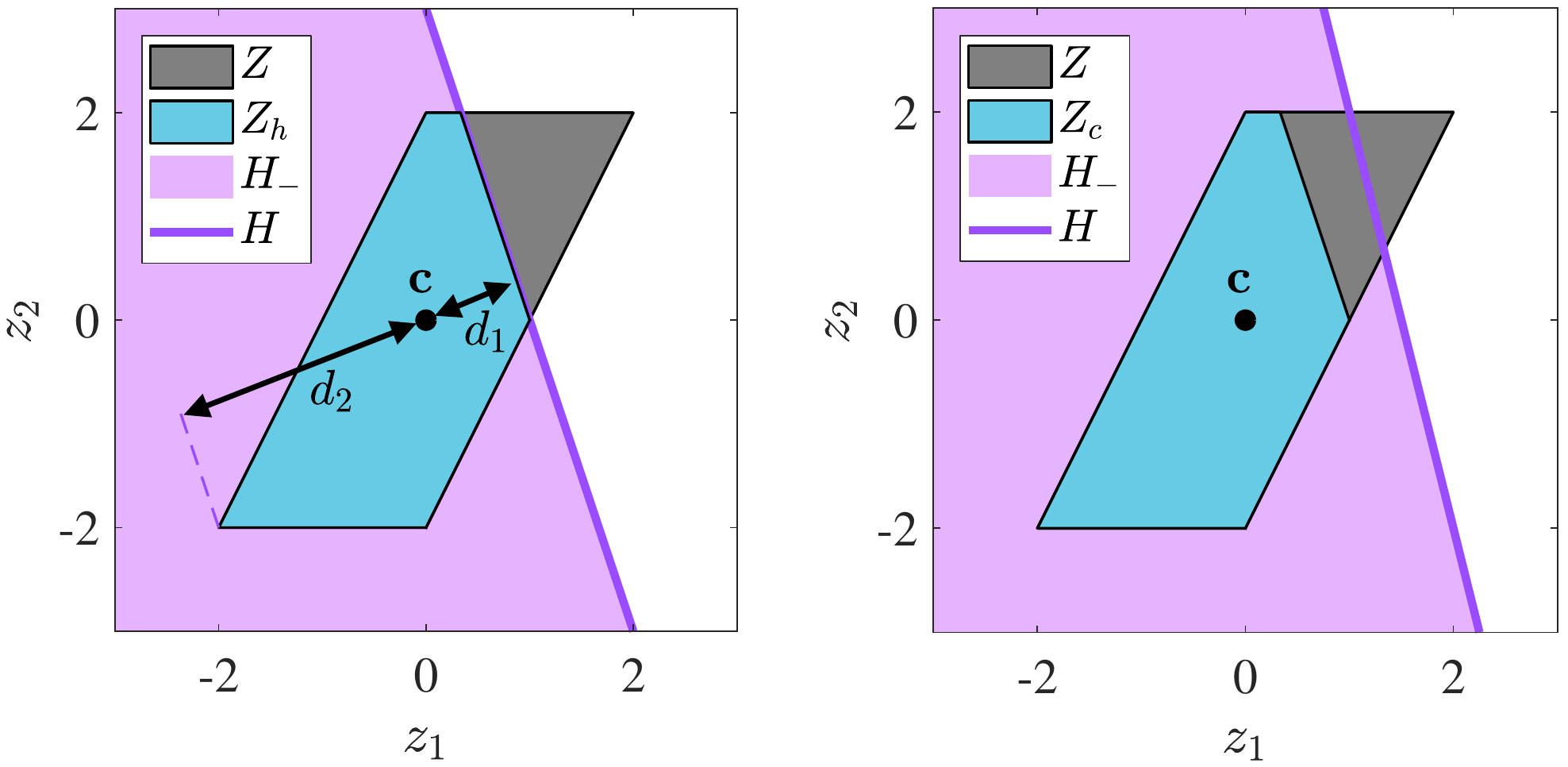}
		\vspace{-10pt}
		\caption{Left: The intersection of the zonotope $ Z $ and the halfspace $ H_- $ corresponding to the hyperplane $ H $ results in the constrained zonotope $ Z_h $. The distances $ d_1 $ and $ d_2 $, measured orthogonally to $ H $, are shown to provide a geometric interpretation of the equality constraints in \eqref{zono_halfspace_int_def}. Right: An example where the constrained zonotope $  Z_c = \{\mathbf{G},\mathbf{c},\mathbf{A},\mathbf{b}\} $, with corresponding unconstrained zonotope  $ Z = \{\mathbf{G},\mathbf{c}\} $, where $ Z $ intersects the hyperplane $ H $ but $ Z_c $ does not.}
		\label{Fig_Halfspace_Intersection}                      
	\end{center}                         
\end{figure}

\begin{thm} \label{zono_halfspace_int}
	If the zonotope $ Z = \{\mathbf{G},\mathbf{c}\} \subset \mathbb{R}^n $ intersects the hyperplane $ H = \{\mathbf{x} \in \mathbb{R}^n \mid \mathbf{h}^T \mathbf{x} = f \} $ corresponding to the halfspace $ H_- = \{\mathbf{x} \in \mathbb{R}^n \mid \mathbf{h}^T \mathbf{x} \leq f \} $, then the intersection $ Z_h = Z \cap H_- $ is a constrained zonotope where
	\begin{equation}\label{zono_halfspace_int_def}
	Z_h = \{[\mathbf{G} \; \mathbf{0}], \mathbf{c}, \begin{bmatrix} \mathbf{h}^T\mathbf{G} \; \frac{d_m}{2}	\end{bmatrix}, \begin{matrix} f - \mathbf{h}^T\mathbf{c}-\frac{d_m}{2}\end{matrix}\}, 
	\end{equation} 
	and $ d_m = f - \mathbf{h}^T\mathbf{c} + \sum_{i = 1}^{n_g} | \mathbf{h}^T \mathbf{g}_i | $.
\end{thm}
\begin{pf}
Considering any element $\mathbf{x} \in Z_h$, it is to be proven that $\mathbf{x} \in Z \cap H_{-}$. From the definition of $Z_h$ in \eqref{zono_halfspace_int_def}, $\exists \; \boldsymbol{\xi} \in \mathbb{R}^{n_g}$ and $\xi_{n_g + 1} \in \mathbb{R}$ such that
\begin{equation*}
    \mathbf{x} = \mathbf{G}\boldsymbol{\xi} + \mathbf{0}\xi_{n_g + 1} + \mathbf{c}, \quad ||\boldsymbol{\xi}||_{\infty} \leq 1, \quad |\xi_{n_g + 1}| \leq 1,
\end{equation*}
\begin{equation}\label{conzono_const}
    \mathbf{h}^T\mathbf{G}\boldsymbol{\xi} + \frac{d_m}{2}\xi_{n_g+1} = f - \mathbf{h}^T\mathbf{c} - \frac{d_m}{2}.
\end{equation}


By the assumption that $Z \cap H \neq \emptyset$, the definition of $d_m$ and \eqref{hyperplane_int} ensure $d_m \geq 0$. If $d_m = 0$, then \eqref{conzono_const} results in $\mathbf{h}^T\mathbf{G}\boldsymbol{\xi} = f- \mathbf{h}^T\mathbf{c}$, which can be rewritten as $\mathbf{h}^T(\mathbf{G}\boldsymbol{\xi} + \mathbf{c}) = f$. Therefore, $\mathbf{x} \in Z_h \subset Z$ and $\mathbf{x} \in H \subset H_{-}$. If $d_m > 0$, \eqref{conzono_const} can be solved for $\xi_{n_g + 1}$ as
\begin{equation} \label{eq_xi_ngplus1}
    \xi_{n_g + 1} = \frac{2}{d_m}(f - \mathbf{h}^T\mathbf{c}- \frac{d_m}{2} - \mathbf{h}^T\mathbf{G}\boldsymbol{\xi}).
\end{equation}
Combining \eqref{eq_xi_ngplus1} and the inequality constraint $-1 \leq \xi_{n_g + 1}$ results in \begin{subequations}
    \begin{align*}
      -1 \leq & \; \xi_{n_g + 1} = \frac{2}{d_m}(f - \mathbf{h}^T\mathbf{c}- \frac{d_m}{2} - \mathbf{h}^T\mathbf{G}\boldsymbol{\xi}),  \\ 
        -\frac{d_m}{2} \leq & \; f - \mathbf{h}^T\mathbf{c} - \frac{d_m}{2} - \mathbf{h}^T\mathbf{G}\boldsymbol{\xi},  \\ 
        \mathbf{h}^T(\mathbf{c} + \mathbf{G}\boldsymbol{\xi}) \leq & \; f. 
    \end{align*}
\end{subequations}
\setcounter{equation}{\value{equation}-1}
Therefore, $\mathbf{x} \in Z$ and $\mathbf{x} \in H_{-}$.
Next, considering any $\mathbf{x} \in Z \cap H_{-}$, it is to be proven that $\mathbf{x} \in Z_h$. For all $\mathbf{x} \in Z \cap H_{-}$, $\exists \; \boldsymbol{\xi} \in \mathbb{R}^{n_g}$ such that 
\begin{equation}\label{eq_x_Z_intsct_H}
\mathbf{x} = \mathbf{G}\boldsymbol{\xi} + \mathbf{c}, \; ||\boldsymbol{\xi}||_{\infty} \leq 1, \; \mathbf{h}^T\mathbf{x} \leq f.
\end{equation}
To show that $\mathbf{x} \in Z_h$ requires proving the existence of $\xi_{n_g + 1} \in \mathbb{R}$ such that
\begin{equation*}
\mathbf{x} = \mathbf{G}\boldsymbol{\xi} + \mathbf{0}\xi_{n_g + 1} + \mathbf{c}, \quad |\xi_{n_g + 1}| \leq 1,
\end{equation*}
and \eqref{conzono_const} holds for all $\mathbf{x}$ satisfying \eqref{eq_x_Z_intsct_H}. If $d_m = 0$, then \eqref{conzono_const} is independent of $\xi_{n_g + 1}$ and holds $\forall \; \mathbf{x} \in Z \cap H_{-}$. Thus, $\xi_{n_g + 1}$ can be arbitrarily chosen such that $|\xi_{n_g + 1}| \leq 1$. If $d_m > 0$, let $\xi_{n_g+1}$ be chosen as in \eqref{eq_xi_ngplus1}, which satisfies \eqref{conzono_const}.
To prove $|\xi_{n_g + 1}| \leq 1$, consider $\mathbf{x}$ as in \eqref{eq_x_Z_intsct_H}. 
Since, $f - \mathbf{h}^T\mathbf{x} \geq 0$, $\xi_{n_g +1}$ satisfies
\begin{equation}\label{xi_ngplus1_exp2}
    \xi_{n_g + 1} = \frac{2}{d_m}(f - \mathbf{h}^T\mathbf{x} - \frac{d_m}{2}) \geq -1.
\end{equation}
Finally, using \eqref{eq_xi_ngplus1}, the fact that $-\mathbf{h}^T\mathbf{G}\boldsymbol{\xi} \leq \sum \limits_{i = 1}^{n_g} |\mathbf{h}^T \mathbf{g}_i |$, and the definition of $d_m$ results in
\begin{subequations}
\begin{align*}
    & \xi_{n_g + 1} \leq \frac{2}{d_m}(f - \mathbf{h}^T\mathbf{c} + \sum\limits_{i = 1}^{n_g}|\mathbf{h}^T \mathbf{g}_i | - \frac{d_m}{2}), \nonumber \\ 
    & \xi_{n_g + 1} \leq \frac{2}{d_m}(d_m - \frac{d_m}{2}) = 1. \nonumber
\end{align*}    
\end{subequations}
\setcounter{equation}{\value{equation}-1}
Thus, $\forall \; \mathbf{x} \in Z \cap H_{-}$, $\mathbf{x} \in Z_h$. \hfill \hfill \qed
 \end{pf}
\begin{exmp}\label{exmp_1}
	The left subplot in Fig. \ref{Fig_Halfspace_Intersection} shows the zonotope $ Z $ and halfspace $ H_- $ where
	\begin{equation*}
	\setlength\arraycolsep{2pt}
	Z = \left\{ \begin{bmatrix} 1 & 1 \\ 0 & 2 \end{bmatrix}, \begin{bmatrix} 0 \\ 0 \end{bmatrix} \right\}, \; H_- = \{\mathbf{x} \in \mathbb{R}^2 \mid \left[ 3 \enspace 1 \right] \mathbf{x} \leq 3 \}.
	\end{equation*}
	From \emph{\textbf{Lemma \ref{Zono_halfspace_int_check}}}, $ Z $ intersects the associated hyperplane $ H $ since \eqref{hyperplane_int} evaluates to $ 3 \leq 8 $. From \emph{\textbf{Theorem \ref{zono_halfspace_int}}}, the intersection $ Z \cap H_- $ is a constrained zonotope and \eqref{zono_halfspace_int_def} evaluates to
	\begin{equation*}
	Z_h = \left\{\begin{bmatrix} 1 & 1 & 0 \\ 0 & 2 & 0 \end{bmatrix}, \begin{bmatrix} 0 \\ 0 \end{bmatrix}, \left[ 3 \enspace 5 \enspace 5.5	\right], -2.5\right\}.
	\end{equation*}
	The left subplot in Fig. \ref{Fig_Halfspace_Intersection} also shows the physical interpretation of $ d_m $ where $ d_m = d_1 + d_2 $.  With $ d_1 = f - \mathbf{h}^T\mathbf{c} $, $ d_1 $ captures the orthogonal distance from the hyperplane $ H $ to the center, $ \mathbf{c} $, of the zonotope. With $ d_2 = \sum_{i = 1}^{n_g} | \mathbf{h}^T \mathbf{g}_i | $, $ d_2 $ captures the orthogonal distance from center of the zonotope to the point in $ Z $ farthest from $ H $.
\end{exmp}

\subsection{Constrained Zonotope-Halfspace Intersection} \label{Sec_conZonoHalfspace}

For the intersection $ Z_h = Z_c \cap H_- $ of a constrained zonotope $ Z_c = \{\mathbf{G},\mathbf{c},\mathbf{A},\mathbf{b}\} $ and a halfspace $ H_- $, \textbf{Theorem~\ref{zono_halfspace_int}} is readily modified where 
\begin{equation} \label{conzono_halfspace_int}
 \scriptsize{Z_h} = \left\{[\mathbf{G} \; \mathbf{0}], \mathbf{c}, \begin{bmatrix} \mathbf{A} & \mathbf{0} \\ \mathbf{h}^T\mathbf{G} & \frac{d_m}{2}	\end{bmatrix}, \begin{bmatrix} \mathbf{b} \\ f - \mathbf{h}^T\mathbf{c}-\frac{d_m}{2}\end{bmatrix}\right\}.
\end{equation} 

However, if the constrained zonotope is completely contained in the halfspace, $ Z_c \subset H_- $, and does not intersect the corresponding hyperplane $ H $, then $ Z_h = Z_c $ and the addition of the $ n_g + 1 $ generator and $ n_c + 1 $ constraint is redundant and increases the order of $ Z_h $ unnecessarily.

However, when determining if a constrained zonotope $ Z_c $ intersects a hyperplane $ H $, the inequality \eqref{hyperplane_int} is necessary but not sufficient. The equality constraints $ \mathbf{A} \boldsymbol{\xi} = \mathbf{b} $ impose restrictions such that $ Z_c \subset Z =  \{\mathbf{G},\mathbf{c}\} $. Thus, the parent zonotope $ Z $ may intersect $ H $ while $ Z_c $ does not (as shown in right subplot of Fig.~\ref{Fig_Halfspace_Intersection}). The intersection of a constrained zonotope with a hyperplane can be checked by solving two Linear Programs (LPs), each with $n_g$ decision variables. 
	
\begin{lem}\label{conzono_hyplane_intersection}
	The constrained zonotope $ Z_c = \{\mathbf{G},\mathbf{c},\mathbf{A},\mathbf{b}\} \subset \mathbb{R}^n $ intersects the hyperplane $ H = \{\mathbf{x} \in \mathbb{R}^n \mid \mathbf{h}^T \mathbf{x} = f \} $ if $ f_{min} \leq f \leq f_{max} $, where
	\begin{subequations}
		\begin{align*}
		f_{min} \triangleq \text{\emph{min}} \{\mathbf{h}^T (\mathbf{c} + \mathbf{G} \boldsymbol{\xi}) \mid \| \boldsymbol{\xi} \|_\infty \leq 1, \mathbf{A} \boldsymbol{\xi} = \mathbf{b} \}, \\
		f_{max} \triangleq \text{\emph{max}} \{\mathbf{h}^T (\mathbf{c} + \mathbf{G} \boldsymbol{\xi}) \mid \| \boldsymbol{\xi} \|_\infty \leq 1, \mathbf{A} \boldsymbol{\xi} = \mathbf{b} \}.
		\end{align*}
	\end{subequations}
\end{lem}
\setcounter{equation}{\value{equation}-1}
\begin{pf}
	From the definition of $ f_{min} $ and $ f_{max} $, if  $ f_{min} \leq f \leq f_{max} $, then there exists $ \mathbf{x}_{min}, \mathbf{x}_{max} \in Z_c $ such that $ \mathbf{h}^T \mathbf{x}_{min} \leq f \leq \mathbf{h}^T \mathbf{x}_{max} $. 
	By the convexity of constrained zonotopes \cite{Scott2016}, there exists $ \mathbf{x}_\lambda \in Z_c $ such that $ \mathbf{x}_\lambda = \lambda \mathbf{x}_{min} + (1-\lambda) \mathbf{x}_{max} $, $ \lambda \in \left[0,1\right] $. For the case where $f_{min} = f_{max} = f $, any choice of $\lambda \in [0,1]$ results in $\mathbf{h}^T\mathbf{x}_{\lambda} = f$. Otherwise, if $f_{min} \neq f_{max}$, choosing $ \lambda = \frac{f-f_{max}}{f_{min}-f_{max}} \in \left[0,1\right] $ results in $ \mathbf{h}^T \mathbf{x}_\lambda = f $. Thus $ \mathbf{x}_\lambda \in H $ and $ \mathbf{x}_\lambda \in Z_c $, proving $ Z_c \cap H \neq \emptyset $. \hfill \hfill \qed
\end{pf}
Note that $f_{min}$ and $f_{max}$ obtained using \textbf{Lemma \ref{conzono_hyplane_intersection}} represent the largest orthogonal distance between a point in $Z_c$ and either side of the hyperplane  providing additional insight to the location of constrained zonotope with respect to the hyperplane.


\begin{rem}\label{rem_conzono_hyp_intersection}
	While the knowledge of $f_{min}$ and $f_{max}$ can be useful, checking for the non-empty intersection of a constrained zonotope and a hyperplane can be achieved by assessing the feasibility of a single LP with constraints
	\begin{equation*}
	    \mathbf{h}^{T}(\mathbf{c} + \mathbf{G}\boldsymbol{\xi}) \leq f, \quad \mathbf{A}\boldsymbol{\xi}= \mathbf{b}, \quad ||\boldsymbol{\xi}||_{\infty} \leq 1.
	\end{equation*}
\end{rem}
When solving these LPs is undesirable, an iterative method based on interval arithmetic from \cite{Scott2016} provides an approach for checking constrained zonotope-halfspace intersection with complexity $O(n_cn_g^2)$. Reproduced from \cite{Scott2016}, \textbf{Algorithm \ref{gen_Bounds}} computes the interval set $ E = [\boldsymbol{\xi}^L, \boldsymbol{\xi}^U] $ such that $ B_\infty(\mathbf{A},\mathbf{b}) \subset E \subset [-\mathbf{1},\mathbf{1}] $ and $ R = [\boldsymbol{\rho}^L, \boldsymbol{\rho}^U] \subset \mathbb{R}^{n_g} $ where
\begin{equation*}
R_j \supset \{\xi_j \mid \mathbf{A} \boldsymbol{\xi} = \mathbf{b}, |\xi_i| \leq 1, \forall i \neq j \}, \quad \forall j \in [1,n_g].
\end{equation*}
As discussed in \cite{Scott2016}, this iterative method has the potential to detect empty constrained zonotopes without solving a LP. Specifically, if $ E \cap R = \emptyset $, then $ Z_c = \emptyset $. Since $ E, R $ are intervals, $ E \cap R = \emptyset $ if $ \xi_j^U < \rho_j^L $ or $ \xi_j^L > \rho_j^U $ for any $ j \in [0,n_g] $.

\IncMargin{1.5em}
\begin{algorithm2e}
	\SetAlgoLined
	\SetKwInOut{Input}{Input}\SetKwInOut{Output}{Output}
	\Input{$ Z_c = \{\mathbf{G},\mathbf{c},\mathbf{A},\mathbf{b}\} $}
	\Output{$ E_j, R_j, \forall j \in [1,n_g] $}
	\BlankLine
	Initialize $ E_j \leftarrow [-1, 1], R_j \leftarrow [-\infty, \infty], \; i,j \leftarrow 1 $
	\While{$ i \leq n_c $}{
		\While{$ j \leq n_g $}{
			\If{$ a_{ij} \neq 0 $}{
				$ R_j \leftarrow R_j \cap ( a_{ij}^{-1}b_i - \sum_{k \neq j} a_{ij}^{-1} a_{ik} E_{k} ) $\;
				$ E_j \leftarrow E_j \cap R_j $\;}
			$ j \leftarrow j+1 $\;}
		$ i \leftarrow i + 1, j \leftarrow 1 $\;}
	\caption{\cite{Scott2016} Constrained zonotope intervals.}
	\label{gen_Bounds}
\end{algorithm2e}
\DecMargin{1.5em}
The goal is to detect if $ Z_c \subset H_- $, resulting in $ Z_h = Z_c $ and thus avoiding the unnecessary addition of generators and constraints from the application of \eqref{conzono_halfspace_int}. The proposed approach uses the fact that $ Z_c \subset H_- $ if and only if $ Z_c \cap H_+ = \emptyset $, where $ H_+ = \{\mathbf{x} \in \mathbb{R}^n \mid \mathbf{h}^T \mathbf{x} \geq f \} $ is the complement of $ H_- $. By modifying \eqref{conzono_halfspace_int} such that $ Z_{h^{+}} = Z_c \cap H_+ $, \textbf{Algorithm \ref{gen_Bounds}} can then be applied to $ Z_{h^{+}} $ to check if $ Z_{h^{+}} = \emptyset $. Specifically, if $ E \cap R = \emptyset $, then $ Z_{h^{+}} = \emptyset $ and $ Z_c \subset H_- $. Note that applying \textbf{Algorithm 1} does not guarantee the detection of $ Z_{h^{+}} = \emptyset $. 
As discussed in \cite{Scott2016}, \textbf{Algorithm~\ref{gen_Bounds}} can be applied iteratively to refine the interval set $ E $. In fact, two iterations of \textbf{Algorithm~\ref{gen_Bounds}} were required to detect that $ Z_c \subset H_- $ for the example shown on the right subplot of Fig. \ref{Fig_Halfspace_Intersection}.

\begin{rem}
To provide an unbiased evaluation of constrained-zonotope hyperplane intersection using \textbf{\emph{Algorithm \ref{gen_Bounds}}}, the intersection of $Z_h$ (from \textbf{\emph{Example \ref{exmp_1}}}) 
with 100 randomly chosen hyperplanes is checked. Note that for all instances, the parent zonotope $Z$ satisfying $Z \supset Z_h$ intersected the random hyperplanes. The constrained zonotope $Z_h$ intersected these random hyperplanes $61$ times and did not intersect for the remaining $39$ times. In all cases, \textbf{\emph{Algorithm \ref{gen_Bounds}}} accurately detected the intersection/non-intersection of the constrained zonotope and randomly generated hyperplanes. Iteration of \textbf{\emph{Algorithm \ref{gen_Bounds}}} to further refine $E$ was only required in $13$ of these $100$ cases.
\end{rem}

\section{Redundancy Removal} \label{Sec_Redundancy}

It is important to recognize that certain set operations can create redundancy in the set representation. For example, the Minkowski sum can create redundancy in the resultant zonotope if the two operands have parallel generators.  Additionally, the generalized intersection can create redundancy within the generators and constraints of a constrained zonotope. Detecting and removing this redundancy can provide order reduction without reducing the volume of the set.
First, if a zonotope $ Z = \{\mathbf{G},\mathbf{c}\} $ has parallel generators, $ \mathbf{g}_i \parallel \mathbf{g}_j $, then the same set can be represented using one less generator by simply combining parallel generators through addition $\mathbf{g}_i + \mathbf{g}_j$. For a zonotope in $ \mathbb{R}^n $ with $ n_g $ generators,  parallel generators can be detected and combined using a typical sorting algorithm with complexity $O(n n_g^2)$. To set a desired numerical precision, two generators are considered parallel if $ \frac{|\mathbf{g}_i^T \mathbf{g}_j|}{\|\mathbf{g}_i\|_2 \|\mathbf{g}_j\|_2} \geq 1 - \epsilon $, where $ \epsilon > 0 $ is a small number. 

The same is true for a constrained zonotope $ Z_c = \{\mathbf{G},\mathbf{c},\mathbf{A},\mathbf{b}\} $ if the lifted zonotope \cite{Scott2016}
\begin{equation*}
Z^{+} = \left\{ \begin{bmatrix} \mathbf{G} \\ \mathbf{A} \end{bmatrix}, \begin{bmatrix} \phantom{-}\mathbf{c} \\ -\mathbf{b} \end{bmatrix} \right\} = \{\mathbf{G}^+,\mathbf{c}^+\},
\end{equation*}
has parallel generators, $ \mathbf{g}_i^+ \parallel \mathbf{g}_j^+ $. In this case, the parallel generators can be similarly reduced but with higher complexity $O(n+n_c)n_g^2$ due to the $ n_c $ constraints added to the rows of the lifted zonotope structure. Once the reduced lifted zonotope is obtained, it
is transformed back to a reduced constrained zonotope with fewer generators.


For constrained zonotopes, redundancy can also come from the combination of constraints $ \mathbf{A} \mathbf{\xi} = \mathbf{b} $ and $ \lVert \boldsymbol{\xi} \rVert_\infty \leq~1 $. By representing these constraints as
\begin{equation} \label{redund_cons}
\mathbf{A} \mathbf{\xi} = \mathbf{b} \Longleftrightarrow \mkern -20mu \sum_{j\in \{ 1, \cdots, n_g \}} \mkern -20mu a_{i,j} \xi_j = b_i, \forall i \in \{ 1, \cdots, n_c\},
\end{equation} 
and $ \lVert \boldsymbol{\xi} \rVert_\infty \leq 1 \Leftrightarrow |\xi_j | \leq 1, \forall j \in \{ 1, \cdots, n_g \}$, the following theorem provides a condition for detecting redundancy and a method for removing one generator and one constraint with complexity $O(n_cn_g^2)$.

\begin{thm} \label{Redundant_ConZono}
	For $ Z_c = \{ \mathbf{G},\mathbf{c},\mathbf{A},\mathbf{b} \} \subset \mathbb{R}^n $ with $ n_g $ generators and $ n_c $ constraints, if there exists indices $ r \in \{ 1, \cdots, n_c \} $ and $ c \in \{1, \cdots, n_g \} $ such that $ a_{r,c} \neq 0 $ and 
	\begin{equation} \label{Ranges_indexed}
	R_{r,c} \triangleq a_{r,c}^{-1} b_r - a_{r,c}^{-1} \sum_{k \neq c} a_{r,k} E_k \subseteq [-1,1],
	\end{equation}
	with $E_k$ computed using \textbf{\emph{Algorithm 1}}, then $ Z_c $ 
	can be exactly represented by a constrained zonotope $ Z_r $ with $ n_g -1 $ generators and $ n_c -1 $ constraints. 
\end{thm}
\begin{pf}
	Following the procedure in \cite{Scott2016}, let  
	\begin{equation*}
	Z_r = \{\mathbf{G}-\mathbf{\Lambda}_G\mathbf{A}, \mathbf{c}+\mathbf{\Lambda}_G\mathbf{b}, \mathbf{A}-\mathbf{\Lambda}_A\mathbf{A}, \mathbf{b}-\mathbf{\Lambda}_A\mathbf{b}\},
	\end{equation*}
	where $ \mathbf{\Lambda}_G = \mathbf{G} \mathbf{E}_{c,r} a_{r,c}^{-1} \in \mathbb{R}^{n \times n_c} $, $ \mathbf{\Lambda}_A = \mathbf{A} \mathbf{E}_{c,r} a_{r,c}^{-1} \in \mathbb{R}^{n_c \times n_c} $, and $ \mathbf{E}_{c,r} \in \mathbb{R}^{n_g \times n_c} $ is zero except for a one in the $ (c,r) $ position. With $ Z_r = \{\mathbf{G}_r,\mathbf{c}_r,\mathbf{A}_r,\mathbf{b}_r\} $, this transformation uses the $ r^{th} $ of row of \eqref{redund_cons} to solve for $ \xi_c $ in terms of $ \xi_k, k\neq c $. This results in the $ c^{th} $ column of $ \mathbf{G}_r $ and $ \mathbf{A}_r $ and the $ r^{th} $ row of $ \mathbf{A}_r $ to equal zero. Removing these columns and rows of zeros results in a constrained zonotope with $ n_g -1 $ generators and $ n_c -1 $ constraints. Through this transformation, the $ r^{th} $ constraint is still imposed in $ Z_r $ but the ability to constraint $ | \xi_c | \leq 1 $ is lost. However, since $ R_{r,c} \subseteq [-1,1] $, this constraint is imposed by the remaining equality and norm constraints, and thus $ Z_r = Z_c $. \hfill \hfill \qed
\end{pf}

\begin{figure}
	\begin{center}
		\includegraphics[width=4cm]{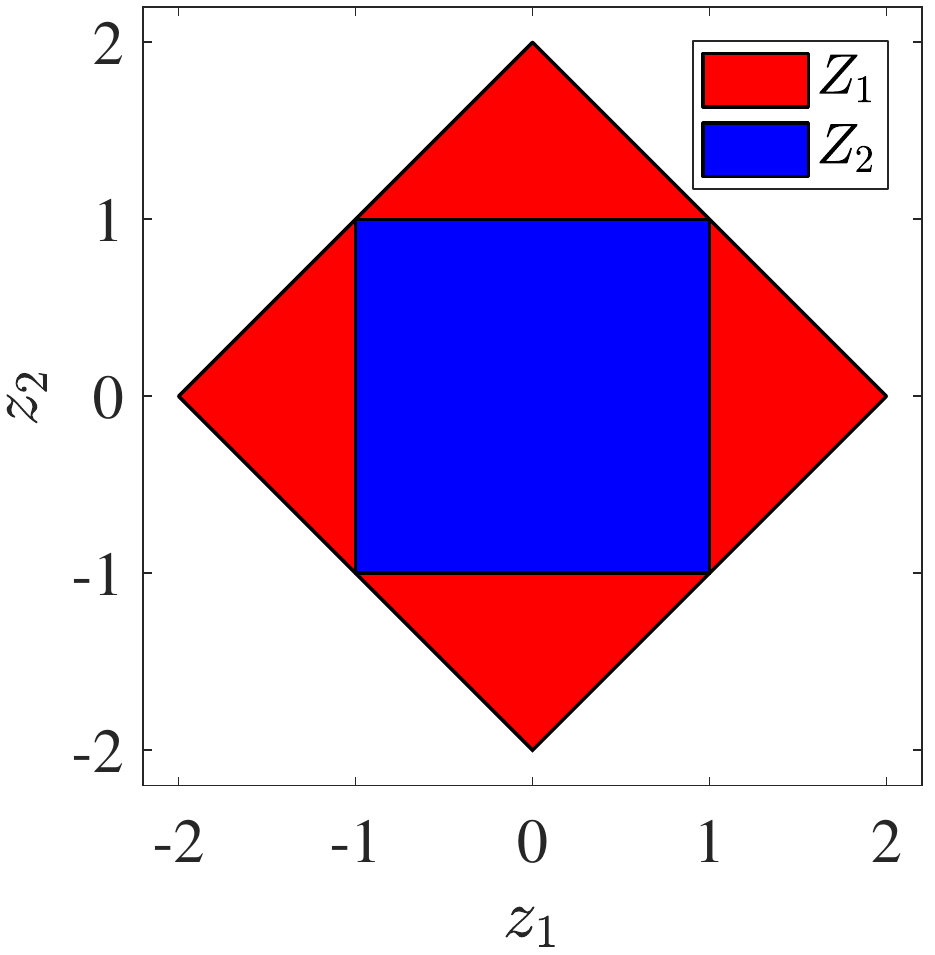}
		\caption{Zonotopes $ Z_1 $ and $ Z_2 $, where $ Z_1 \cap Z_2 = Z_2 $, used to demonstrate the ability to remove redundancy from constrained zonotopes that can arise from operations like the generalized intersection.}
		\label{Fig_Redundancy}                 	\end{center}                        
\end{figure}

As in \cite{Scott2016}, Gauss-Jordan elimination with full pivoting should be applied to $ Z_c $ prior to applying \textbf{Algorithm \ref{gen_Bounds}} to determine the intervals $  E_k $ required to compute \eqref{Ranges_indexed}.
The procedure discussed in the proof of \textbf{Theorem \ref{Redundant_ConZono}} can be applied iteratively until $ R_{r,c} \nsubseteq [-1,1] $ for any indices. However, there is no guarantee that the resulting constrained zonotope will be without redundancy since \textbf{Theorem \ref{Redundant_ConZono}} only provides a sufficient condition. 
\begin{exmp}
	Consider the two zonotopes shown in Fig. \emph{\ref{Fig_Redundancy}}
	\begin{equation*}
	Z_1 = \left\{ \begin{bmatrix} 1 & \phantom{-}1 \\ 1 & -1 \end{bmatrix}, \begin{bmatrix} 0 \\ 0 \end{bmatrix} \right\}, \quad
	Z_2 = \left\{ \begin{bmatrix} 1 & 0 \\ 0 & 1 \end{bmatrix}, \begin{bmatrix} 0 \\ 0 \end{bmatrix} \right\},
	\end{equation*}
 and the constrained zonotope $ Z_c = Z_1 \cap Z_2 $.  Applying \eqref{generalized_Intersection} results in 
	\begin{equation} \label{Redund_gen_int_example}
	\setlength\arraycolsep{1.7pt}
	Z_c = \left\{ \begin{bmatrix} 1 & \phantom{-}1 & 0 & 0 \\ 1 & -1 & 0 & 0 \end{bmatrix}, \begin{bmatrix} 0 \\ 0 \end{bmatrix}, \begin{bmatrix} 1 & \phantom{-}1 & -1 & \phantom{-}0 \\ 1 & -1 & \phantom{-}0 & -1 \end{bmatrix}, \begin{bmatrix} 0 \\ 0 \end{bmatrix} \right\},
	\end{equation}
	with $ n_g = 4 $ generators and $ n_c = 2 $ constraints. However, since $ Z_2 \subset Z_1 $, the intersection is also represented exactly by $ Z_2 $. By applying Gauss-Jordan elimination with full pivoting and two iterations of the procedure from \emph{\textbf{Theorem \ref{Redundant_ConZono}}}, two constraints and two generators are removed to reduce $ Z_c $ from \eqref{Redund_gen_int_example} to $ Z_c = Z_2 $ with $ n_g = 2 $ and $ n_c = 0 $. To provide an unbiased evaluation of \emph{\textbf{Theorem \ref{Redundant_ConZono}}}, the axis-aligned generators of $Z_2$ above were replaced by randomly chosen generators. In each of the $45$ out of $100$ cases where $Z_2 \subseteq Z_1$, $Z_c$ was successfully reduced to $Z_c = Z_2$ with $n_g = 2$ and $n_c = 0$. 
	\end{exmp}

\begin{rem}
For a constrained zonotope $Z_c$ with $n_c$ constraints and $n_g$ generators and a set $H$ in H-Rep with with $n_h$ halfspaces, \textbf{\emph{Algorithm \ref{gen_Bounds}}} can be applied in two different ways to either prevent or remove redundancy in the set representation of $ Z_c \cap H $. The approach from Section \ref{Sec_conZonoHalfspace} based on preventing the addition of unnecessary generators and constraints has a best-case complexity of $O(n_h n_c n_g^2)$ if $ Z_c \subset H $ and a worst-case complexity of $O(n_h (n_c+n_h) (n_g+n_h)^2)$ if $ Z_c $ intersects each of the $ n_h $ halfspaces. Alternatively, $ n_h $ constraints and $n_h$ generators can be directly added to $ Z_c $ using \eqref{conzono_halfspace_int} and then \textbf{\emph{Theorem \ref{Redundant_ConZono}}} can be applied to reduce set complexity. This approach has a best-case complexity of $O((n_c+n_h)(n_g+n_h)^2)$ when no generators/constraints can be removed and a worst-case complexity $O(n_h(n_c+n_h)(n_g+n_h)^2)$ when all of the added $ n_h $ constraints and $n_h$ generators can be removed. Thus, both approaches have the same worst-case complexity but the preventative approach has the potential to require fewer computations in practice.
\end{rem}

\section{Inner-Approximations} \label{Sec_InnerApprox}

Once attempts have been made to remove redundancy from the representation of a zonotope or constrained zonotope, further complexity reduction may be required. As discussed in the Introduction, the majority of order reduction techniques have focused on outer-approximations. This section establishes inner-approximation order reduction for zonotopes and constrained zonotopes.

\subsection{Zonotopes}

\begin{figure}
	\begin{center}
		\includegraphics[width=4cm]{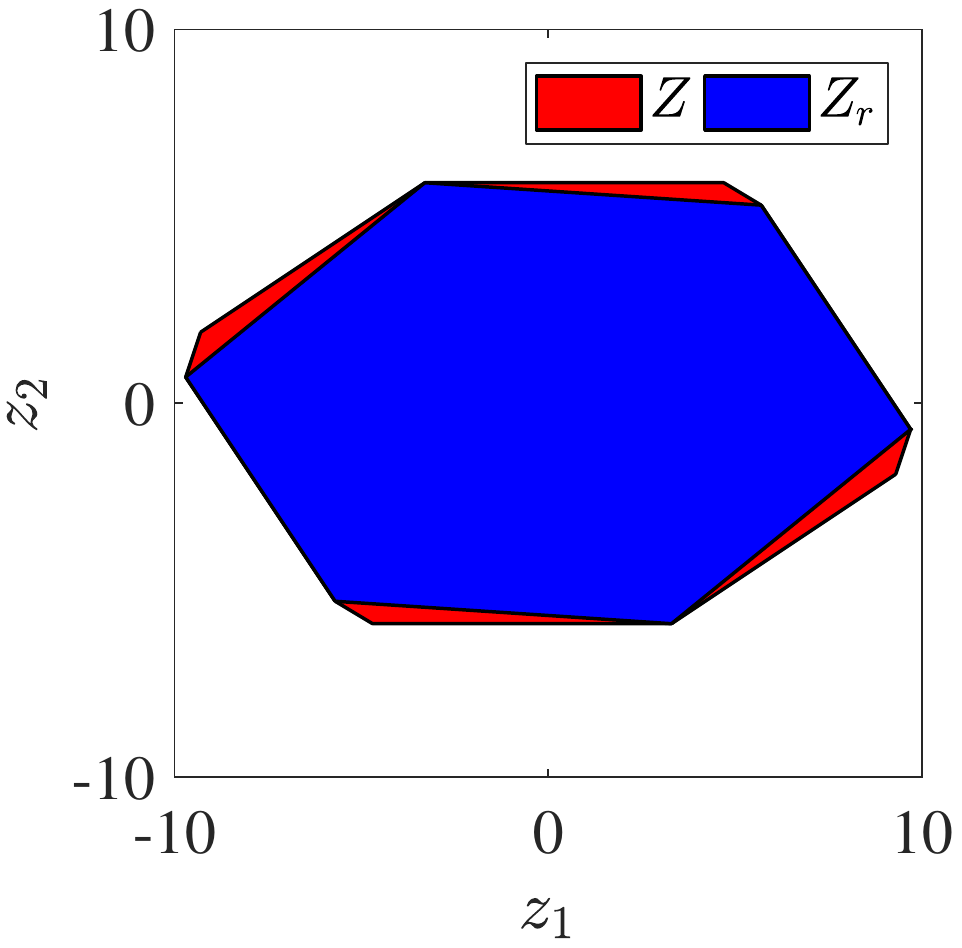}
		\caption{The inner-approximation of $ Z $ with $ n_g = 5 $ by the reduced-order zonotope $ Z_r $ with $ n_r = 3 $.}
		\label{Fig_Zono_Inapprox_Tmat}              
	\end{center}                         
\end{figure}

The proposed reduced-order inner-approximation of a zonotope requires the following zonotope containment conditions.
\begin{lem} (Theorem 3 of
	\emph{\cite{Sadraddini2019}}) \label{zonoContainment}
	Given two zonotopes $ X = \{\mathbf{G}_x,\mathbf{c}_x\} \subset \mathbb{R}^n $ and $ Y = \{\mathbf{G}_y,\mathbf{c}_y\} \subset \mathbb{R}^n $, $ X \subseteq Y $ if there exists $ \mathbf{\Gamma} \in \mathbb{R}^{n_y \times n_x} $ and $ \boldsymbol{\beta} \in \mathbb{R}^{n_y} $ such that
	\begin{equation} \label{zonoContainment_Conditions}
	\mathbf{G}_x = \mathbf{G}_y \mathbf{\Gamma}, \quad \mathbf{c}_y - \mathbf{c}_x = \mathbf{G}_y \boldsymbol{\beta}, \quad |\mathbf{\Gamma}|\mathbf{1} + |\boldsymbol{\beta}| \leq \mathbf{1}.
	\end{equation}
\end{lem}
\begin{thm} \label{ZonoInnerApprox}
	The zonotope $ Z_r = \{\mathbf{G}_r,\mathbf{c}\} \subset \mathbb{R}^n $ is a reduced-order inner-approximation of $ Z = \{\mathbf{G},\mathbf{c}\} \subset \mathbb{R}^n $ such that $ Z_r \subseteq Z $ with $ \mathbf{G}_r \in \mathbb{R}^{n \times n_r} $, $ \mathbf{G} \in \mathbb{R}^{n \times n_g} $, and $ n_r < n_g $ if $ \mathbf{G}_r = \mathbf{G} \mathbf{T} $ where $ \mathbf{T} = [t_{i,j}] \in \mathbb{R}^{n_g \times n_r} $, $ t_{i,j} \in \{-1,0,1\} $, and $ \sum_{j=1}^{n_r} |t_{i,j}| = 1 $, $\forall i \in \{ 1, \cdots, n_g\}$.
\end{thm}
\begin{pf}
	From \textbf{Lemma \ref{zonoContainment}}, $ Z_r \subseteq Z $ if there exist $ \mathbf{\Gamma} \in \mathbb{R}^{n_g \times n_r} $ and $ \boldsymbol{\beta} \in \mathbb{R}^{n_g} $ such that
	\begin{equation*}
	\mathbf{G}\mathbf{T} = \mathbf{G} \mathbf{\Gamma}, \quad \mathbf{c} - \mathbf{c} = \mathbf{G} \boldsymbol{\beta}, \quad |\mathbf{\Gamma}|\mathbf{1} + |\boldsymbol{\beta}| \leq \mathbf{1}.
	\end{equation*}
	The first two equations hold by setting $ \mathbf{\Gamma} = \mathbf{T} $ and $ \boldsymbol{\beta} = \mathbf{0} $. The third equation holds since $ \sum_{j=1}^{n_r} |t_{i,j}| = 1, \; \forall i \in \{1, \cdots, n_g\} $, if and only if $ |\mathbf{T}|\mathbf{1} = \mathbf{1} $. \hfill \hfill \qed
\end{pf}
The specific definition of $ \mathbf{T} $ in \textbf{Theorem \ref{ZonoInnerApprox}} produces an inner-approximation of $ Z $ by forming the generators of $ Z_r $ through the addition of the generators in $ Z $. Typically, the largest inner-approximation of $ Z $ is desired. The proposed method for determining $ \mathbf{T} $ is inspired by the methods for determining outer-approximations of zonotopes presented in \cite{Kopetzki2017a}. First, let the generators $ \mathbf{g}_i $ of $ Z $ be arranged such that $ \| \mathbf{g}_i \|_2 \geq \| \mathbf{g}_{i+1} \|_2, \; \forall i \in \{1, \cdots, n_g -1 \} $. Then partition the generator matrix such that $ \mathbf{G} = [\mathbf{G}_1 \; \mathbf{G}_2] $ where $ \mathbf{G}_1 \in \mathbb{R}^{n \times n_r} $ and $ \mathbf{G}_2 \in \mathbb{R}^{n \times (n_g-n_r)} $. For each generator $ \mathbf{g}_{2,j} $ in $ \mathbf{G}_2 $, compute the magnitude of the dot product $ \alpha_{i,j} = | \mathbf{g}_{1,i}^T \mathbf{g}_{2,j}| $ with all generators $ \mathbf{g}_{1,i} $ in $ \mathbf{G}_1 $. The goal is to add the generators $ \mathbf{g}_{2,j} $ to the most aligned generator $ \mathbf{g}_{1,i} $. Thus, let $ \mathbf{T} = [t_{i,j}] $ where
\begin{equation} \label{T_zonoapprox}
t_{i,j} = \begin{Bmatrix} 1 & \text{if } i=j\leq n_r \\
\frac{1}{\alpha_{i,j}}\mathbf{g}_{1,i}^T \mathbf{g}_{2,j} & \text{if } \alpha_{i,j} > \alpha_{i,k}, \forall k \neq j \\
0 & \text{otherwise}
\end{Bmatrix}.
\end{equation}
Note that computing $Z_r$ using \textbf{Theorem \ref{ZonoInnerApprox}} and \eqref{T_zonoapprox} has an overall complexity of $O(nn_g^2 + nn_gn_r)$, where the first term is associated with sorting the generators based on the $2$-norm and the second term is associated with computing the product $\mathbf{G}_r = \mathbf{G}\mathbf{T}$ in \textbf{Theorem \ref{ZonoInnerApprox}}.
\begin{exmp}\label{exmp_zon_innerapprox}
	Consider the zonotope 
	\begin{equation*}
	Z = \left\{ \begin{bmatrix} 4 & 3 & -2 & 0.2 & 0.5 \\ 0 & 2 & 3 & 0.6 & -0.3 \end{bmatrix}, \boldsymbol{0} \right\} \subset \mathbb{R}^2. 
	\end{equation*}
	Note that the generators are already arranged in order of decreasing 2-norm. With $ n_g = 5 $, the goal is to determine $ Z_r \subseteq Z $ such that $ n_r = 3 $. From \textbf{\emph{Theorem \ref{ZonoInnerApprox}}} and \eqref{T_zonoapprox}, the matrix $\mathbf{T}$ and the reduced-order zonotope $Z_r$ are 
	\begin{equation*}
	    \mathbf{T} = \begin{bmatrix} 1 & 0 & 0 \\ 0 & 1 & 0 \\ 0 & 0 & 1 \\ 0 & 1 & 0 \\ 1 & 0 & 0 \end{bmatrix}, \; Z_r = \left\{ \begin{bmatrix} \phantom{-}4.5 & 3.2 & -2 \\ -0.3 & 2.6 & \phantom{-}3 \end{bmatrix}, \boldsymbol{0} \right\}.
	\end{equation*}

	Fig. \emph{\ref{Fig_Zono_Inapprox_Tmat}} confirms $ Z_r \subseteq Z $ with volume ratio $V_r = 0.97$. 
	While this numerical example resulted in relatively large volume ratio, the reduction in volume is highly dependent on the distribution of generator lengths and the number of generators removed. For 100 randomly generated zonotopes in $ \mathbb{R}^2 $ with $n_g = 5$, applying \textbf{\emph{Theorem \ref{ZonoInnerApprox}}} and \eqref{T_zonoapprox}, resulted in all reduced zonotopes satisfying $Z_r \subseteq Z$ with $n_r = 3$ and mean volume ratio $V_r = 0.84$.
\end{exmp}

\subsection{Constrained Zonotopes}\label{subsec_conzono_innerapprox}

For constrained zonotopes, a reduced-order inner-approximation $ Z_r $ of $ Z_c $ can be computed based on the set containment criteria for the affine transformation of polytopes in H-Rep (AH-polytopes) developed in \cite{Sadraddini2019} since AH-polytopes and constrained zonotopes are equivalent.

\begin{defn} \label{AHPolytope_def} \emph{\cite{Sadraddini2019}}
	An AH-polytope $ X \subset \mathbb{R}^n$ is an affine transformation of a H-Rep polytope $ P \subset \mathbb{R}^m$ where
	\begin{equation} \label{AH_def}
	X = \bar{\mathbf{x}} + \mathbf{X} P, \quad \mathbf{X} \in \mathbb{R}^{n \times m}, \quad \bar{\mathbf{x}} \in \mathbb{R}^n.
	\end{equation}
\end{defn}

The following theorem proves the equivalency between constrained zonotopes and AH-polytopes in addition to providing a method to convert constrained zonotopes to AH-polytopes with complexity $O(nn_g^2+n_c^2n_g)$,
where the first term is associated with computing an affine transformation and the second term is associated with computing the basis of $\mathcal{N}(\mathbf{A})$ for $Z_c = \{\mathbf{G}, \mathbf{c}, \mathbf{A}, \mathbf{b} \}$.
	\begin{thm} \label{ConZono_AHpolytope}
	A non-empty set $ Z_c \subset \mathbb{R}^n $ is a constrained zonotope if and only if it is an AH-polytope.
\end{thm}
\begin{pf}
	To prove that every AH-polytope is a constrained zonotope, let $ P = \{ \mathbf{z} \in \mathbb{R}^m \mid \mathbf{H} \mathbf{z} \leq \mathbf{k} \} $. Per Theorem 1 in \cite{Scott2016}, the set $ P $ can always be represented as a constrained zonotope $ P =  \{\mathbf{G}_p, \mathbf{c}_p, \mathbf{A}_p, \mathbf{b}_p\} $. Thus, from \eqref{AH_def} and the properties of constrained zonotopes \eqref{affineMap} and \eqref{MinkowskiSum}, $ X $ is a constrained zonotope where $ X = \{\mathbf{X}\mathbf{G}_p, \bar{\mathbf{x}}+\mathbf{X}\mathbf{c}_p, \mathbf{A}_p, \mathbf{b}_p\} $. 
	To prove that every constrained zonotope is an AH-polytope, consider $ Z_c  =  \{\mathbf{G}, \mathbf{c}, \mathbf{A}, \mathbf{b}\} $ with $ n_g $ generators and $ n_c $ constraints. If $ n_c = 0 $, $ Z_c = Z = \{\mathbf{G},\mathbf{c}\} $ is a zonotope and can be represented in AH-polytope form of \eqref{AH_def} with $ \bar{\mathbf{x}} = \mathbf{c} $, $ \mathbf{X} = \mathbf{G} $, and $ P = B_\infty $. For $ n_c > 0 $, assume that any rank deficiency in $ \mathbf{A} $ has been detected as a row of zeros in the reduced row echelon form achieved through Gauss-Jordan elimination with full pivoting (see \cite{Scott2016} for details). Thus, the rank of $ \mathbf{A} $ is $ n_c $ and there exists $\mathbf{s} = \mathbf{A}^{\dagger}\mathbf{b} \in \mathbb{R}^{n_g}$ and the matrix $ \mathbf{T} \in \mathbb{R}^{n_g \times (n_g-n_c)} $ with columns that form a basis for $\mathcal{N}(\mathbf{A})$. Using the change of variables $ \boldsymbol{\xi} = \mathbf{T} \bar{\boldsymbol{\xi}} + \mathbf{s} $, the equality constraint $\mathbf{A}\boldsymbol{\xi} = \mathbf{b}$\\ is satisfied for all $ \bar{\boldsymbol{\xi}} \in \mathbb{R}^{n_g-n_c} $.
    Hence, $Z_c$ can be expressed as 
	\begin{equation*}
	Z_c = \left\{ \mathbf{c}+\mathbf{G}\mathbf{s} + \mathbf{G}\mathbf{T}\bar{\boldsymbol{\xi}} \mid \| \mathbf{T} \bar{\boldsymbol{\xi}} + \mathbf{s} \|_\infty \leq 1 \right\}.
	\end{equation*}
	Furthermore, the norm constraints $ \| \mathbf{T} \bar{\boldsymbol{\xi}} + \mathbf{s} \|_\infty \leq 1 $ can be represented in H-Rep as $P = \{ \bar{\boldsymbol{\xi}} \mid \mathbf{H} \bar{\boldsymbol{\xi}} \leq \mathbf{k} \}$, where 
	\begin{equation*}
	\mathbf{H} = \begin{bmatrix} \phantom{-}\mathbf{T} \\ -\mathbf{T} \end{bmatrix},  \quad 
	\mathbf{k} = \begin{bmatrix} \mathbf{1} - \mathbf{s} \\ \mathbf{1} + \mathbf{s} \end{bmatrix}.
	\end{equation*}
	Thus, with $ \bar{\mathbf{x}} = \mathbf{c}+\mathbf{G}\mathbf{s} $ and $ \mathbf{X} =  \mathbf{G}\mathbf{T} $,
$ Z_c $ is an AH-polytope of the form \eqref{AH_def}. 	\hfill \hfill \qed
\end{pf}


\begin{rem}
		The convexity of the constrained zonotope $Z_c = (\mathbf{G}, \mathbf{c}, \mathbf{A}, \mathbf{b})$ also facilitates representation as a polynomial zonotope $Z_p = (\mathbf{c}, \mathbf{G}, \mathbf{E})$ in Z-Rep \cite{Kochdumper2019}. However, the reverse is not true.
\end{rem}
\begin{lem}(Theorem 1 of \emph{\cite{Sadraddini2019}}) \label{AHContainment}
	Given AH-polytopes $ X,Y \subset \mathbb{R}^n $ where $ X = \bar{\mathbf{x}} + \mathbf{X} P_x $, $ Y = \bar{\mathbf{y}} + \mathbf{Y} P_y $, $ P_x = \left\{ \mathbf{x} \in \mathbb{R}^{n_x} \mid \mathbf{H}_x \mathbf{x} \leq \mathbf{f}_x \right\} $, and $ P_y = \left\{ \mathbf{y} \in \mathbb{R}^{n_y} \mid \mathbf{H}_y \mathbf{y} \leq \mathbf{f}_y \right\}$, $ X \subseteq Y $ if there exists $\mathbf{\Gamma} \in \mathbb{R}^{n_y \times n_x}, \boldsymbol{\beta} \in \mathbb{R}^{n_y}$ and $\mathbf{\Lambda} \in \mathbb{R}^{n_{hy} \times n_{hx}}_{+}$ such that 
	\begin{subequations} \label{AHContainment_Cons}
		\begin{align}
		\mathbf{X} &= \mathbf{Y} \mathbf{\Gamma}, & \bar{\mathbf{y}} - \bar{\mathbf{x}} &= \mathbf{Y} \boldsymbol{\beta}, \\ 
		\mathbf{\Lambda} \mathbf{H}_x &= \mathbf{H}_y \mathbf{\Gamma}, & \mathbf{\Lambda} \mathbf{f}_x &\leq \mathbf{f}_y + \mathbf{H}_y \boldsymbol{\beta}. 
		\end{align}
	\end{subequations}
\end{lem}

To achieve a reduced-order inner-approximation $ Z_r $ of constrained zonotope $ Z_c $, \textbf{Theorem \ref{ConZono_AHpolytope}} can be used to convert both $ Z_r $ and $ Z_c $ in to AH-polytopes while \textbf{Lemma \ref{AHContainment}} can be used to ensure $ Z_r \subseteq Z_c $. Assuming $ Z_c $ is known, consider $ Z_r =  \{\mathbf{G}_r \mathbf{\Phi}, \mathbf{c}_r, \mathbf{A}_r, \mathbf{b}_r\} $ where $ \mathbf{\Phi} = diag(\boldsymbol{\phi}) $ is a scaling matrix with $ \phi_i > 0, \forall i \in \{1,\cdots, n_{gr} \} $. Assuming $ \mathbf{G}_r $, $ \mathbf{A}_r $, and $ \mathbf{b}_r $ are known, the following optimization problem can be formulated with $4n_{gr}^2 + n_{gr} + (n_g - n_c)(1 + n_{gr} - n_{cr}) + n$ decision variables that maximizes the $p = 1$, $2$, or $\infty$ norm of the diagonal elements $\phi$ of the scaling matrix $\mathbf{\Phi}$ by solving
\begin{subequations}\label{Conzono_containment}
	\begin{align}
	& \underset{\mathbf{\Phi}, \mathbf{\Gamma}, \boldsymbol{\beta}, \mathbf{\Lambda}, \mathbf{c}_r}{\text{max}} \mkern 20mu ||
	\mathbf{\phi}
	||_p, \\
	& \text{s.t.} \nonumber \\
	& (\mathbf{c} + \mathbf{G}\mathbf{s}) - (\mathbf{c}_r + \mathbf{G}_r\mathbf{\Phi}\mathbf{s}_r) = \mathbf{G}\mathbf{T}\boldsymbol{\beta}, \\
	& \mathbf{G}_r\mathbf{\Phi}\mathbf{T}_r = \mathbf{G}\mathbf{T}\mathbf{\Gamma},\quad  \mathbf{\Lambda}\begin{bmatrix}\phantom{-}\mathbf{T}_r \\ \mathbf{-T}_r\end{bmatrix} = \begin{bmatrix} \phantom{-}\mathbf{T} \\ -\mathbf{T} \end{bmatrix}\mathbf{\Gamma}, \\
	& \mathbf{\Lambda}\begin{bmatrix} \mathbf{1} - \mathbf{s}_r \\ \mathbf{1} + \mathbf{s}_r \end{bmatrix} \leq \begin{bmatrix} \mathbf{1} - \mathbf{s} \\ \mathbf{1} + \mathbf{s} \end{bmatrix} + \begin{bmatrix} \phantom{-}\mathbf{T} \\ -\mathbf{T}\end{bmatrix}\boldsymbol{\beta},
	\end{align}
\end{subequations}
with parameters $\mathbf{s} = \mathbf{A}^{\dagger}\mathbf{b} \in \mathbb{R}^{n_g} $, $ \mathbf{s}_r = \mathbf{A}_r^{\dagger}\mathbf{b}_r \in \mathbb{R}^{n_{gr}} $, and matrices $ \mathbf{T} \in \mathbb{R}^{n_g \times (n_g - n_c)}, \mathbf{T}_r \in \mathbb{R}^{n_{gr} \times (n_{gr} - n_{cr})} $ with columns that form bases for $ \mathcal{N}(\mathbf{A}) $ and $ \mathcal{N}(\mathbf{A}_r) $, respectively.
Note that the majority of the decision variables in \eqref{Conzono_containment} come from the matrices $\mathbf{\Gamma} \in \mathbb{R}^{(n_g - n_c) \times (n_{gr} - n_{cr})}$ and $\mathbf{\Lambda} \in \mathbb{R}^{2n_{gr} \times 2n_{gr}}_{+}$. 
While this procedure applies to any $ Z_r $, the process discussed in Section \ref{Sec_Redundancy} can be used to compute $ Z_r $ by removing exactly one constraint and one generator from $ Z_c $. For the case where $ Z_c $ satisfies the conditions in \textbf{Theorem~\ref{Redundant_ConZono}}, the $ r^{th} $ constraint and the $ c^{th} $ generators were chosen such that $ R_{r,c} \subseteq [-1,1] $ and thus an exact reduced-order representation was achieved with $ Z_r = Z_c $. To achieve further reduction through the inner-approximation of $ Z_c $, the same procedure from Section \ref{Sec_Redundancy} can be applied by choosing appropriate indices and scaling $ Z_r $ via optimization while enforcing $ Z_r \subseteq Z_c $ using the constraints from \eqref{Conzono_containment}. 
Since $ R_j = [\rho_j^L, \rho_j^U] $ represents the range of $ \xi_j $ if the constraints $ | \xi_j | \leq 1 $ were omitted \cite{Scott2016}, the $ c^{th} $ generator should be removed that minimizes $ \max(|\rho_j^L|, |\rho_j^U|) $. Once $ c $ is chosen, $ r $ should be chosen such that the entry in the $ (r,c) $ position of $ \mathbf{A}_r $ has the largest absolute value of all entries in the $ c^{th} $ column. 

\begin{figure}
	\begin{center}
		\includegraphics[width=8.6cm]{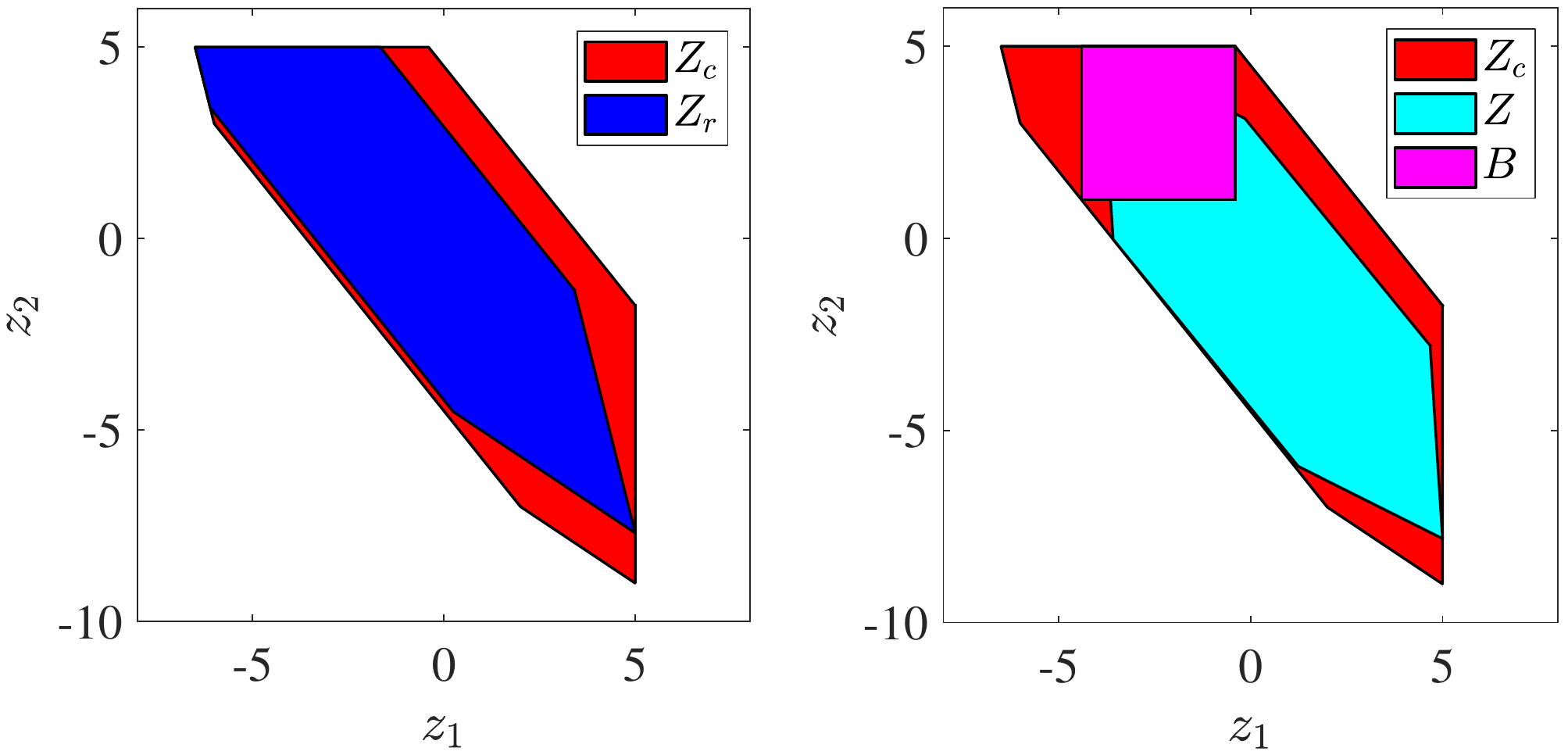}
		\caption{Left: The inner-approximation of $ Z_c $ by a constrained zonotope $ Z_r $ with one less generator and constraint. Right: The inner-approximation of $ Z_c $ by a zonotope $ Z $ and an interval set $ B $.}
		\label{Fig_ConZono_Inapprox}             
	\end{center}      
\end{figure}

\begin{figure}
	\begin{center}
		\includegraphics[width=5cm]{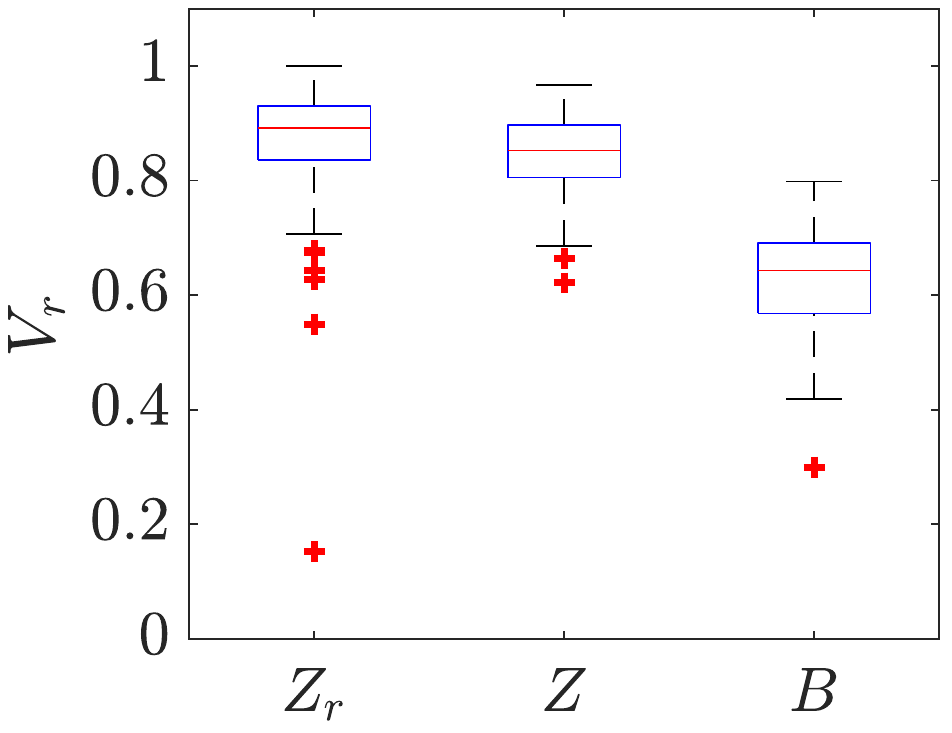}
		\caption{The volume ratios for the inner-approximation of 100 randomly generated constrained zonotopes by a constrained zonotope $ Z_r $ with one less generator and constraint, a zonotope $ Z $, and an interval set $ B $. The red crosses denote outliers that do not fit the box plot distribution.}
		\label{Fig_ConZono_Inapprox_Rand} 	\end{center}               
\end{figure}

\begin{exmp} \label{example_innerApprox}
	Consider the constrained zonotope $ Z_c $ shown in Fig. \ref{Fig_ConZono_Inapprox} where
	\begin{subequations}
		\setlength\arraycolsep{2pt}
		\begin{align*}
		Z_c = \Bigg \{ &\begin{bmatrix} -1 & \phantom{-}3 & \phantom{-}4 & 0 & 0 \\ \phantom{-}4 & -2 & -5 & 0 & 0 \end{bmatrix},  \begin{bmatrix} 0 \\ 0 \end{bmatrix}, \\
		&\begin{bmatrix} -1 & \phantom{-}3 & \phantom{-}4 & 6.5 & 0 \\ \phantom{-}4 & -2 & -5 & 0 & 8 \end{bmatrix}, \begin{bmatrix} -1.5 \\ -3 \end{bmatrix} \Bigg \}.
		\end{align*}
	\end{subequations}
	\setcounter{equation}{\value{equation}-1}
	First, Gauss-Jordan elimination with full pivoting was applied to $ Z_c $, followed by the transformation in \emph{\textbf{Theorem \ref{Redundant_ConZono}}} by picking the $ c^{th} $ generator that minimizes $ \max(|\rho_j^L|, |\rho_j^U|) $ and the $ r^{th} $ row with the largest entry in $ c^{th} $ column of $ \mathbf{A} $. Then an LP was formulated and solved using the constraints from \eqref{AHContainment_Cons} and a cost function that maximized $ \| \boldsymbol{\phi} \|_\infty $. The resulting reduced-order zonotope $ Z_r $ is shown in Fig. \ref{Fig_ConZono_Inapprox} where 
	\begin{subequations}
		\setlength\arraycolsep{2pt}
		\begin{align*}
		Z_r = \Bigg \{ &\begin{bmatrix} 0 & \phantom{-}3.17 & \phantom{-}2.38 & -0.79 \\ 0 & -3.97 & -1.59 & \phantom{-}3.17 \end{bmatrix},  \begin{bmatrix} -1.34 \\ \phantom{-}1.03 \end{bmatrix}, \\
		&\begin{bmatrix} 1 & -0.63 & -0.25 & \phantom{-}0.50 \end{bmatrix}, \begin{bmatrix} -0.38 \end{bmatrix} \Bigg \}.
		\end{align*}
	\end{subequations}
	\setcounter{equation}{\value{equation}-1}
Using a similar approach, Fig, \ref{Fig_ConZono_Inapprox} also shows the inner-approximations of $ Z_c $ by zonotope $ Z $ and interval set $ B $ where
	\begin{subequations}
		\setlength\arraycolsep{2pt}
		\begin{align*}
		Z &= \left\{ \begin{bmatrix} \phantom{-}2.31 & \phantom{-}1.93 & -0.27 \\ -2.84 & -0.94 & \phantom{-} 2.57\end{bmatrix}, \begin{bmatrix} \phantom{-}0.49 \\ -1.35 \end{bmatrix} \right\}, \\
		B &= \left\{ \begin{bmatrix} 2 & 0 \\ 0 & 2 \end{bmatrix}, \begin{bmatrix} \phantom{-}2.55 \\ -3.18 \end{bmatrix} \right\}.
		\end{align*}
	\end{subequations}
	\setcounter{equation}{\value{equation}-1}
	To compute $ Z $, the equality constraints from $ Z_c $ were removed via the same change of variables used in the proof of \emph{\textbf{Theorem~\ref{ConZono_AHpolytope}}}.
	Typically this would result in an outer-approximation of $ Z_c $, however the scaling matrix $ \mathbf{\Phi} $ is used to reduce the length of each generator such that $ Z \subseteq Z_c $. For the interval set $ B $, the generator matrix is initialized as the identity matrix and then scaled by $ \mathbf{\Phi} $. The resulting volume ratios with respect to $ Z_c $ are $ V_r = 0.86 $, $ V_r = 0.83 $, $ V_r = 0.46 $ for $ Z_r $, $ Z $, and $ B $, respectively.  
	Repeating this process for 100 randomly generated constrained zonotopes with $ 4 \leq n_g \leq 20 $ and $ 1 \leq n_c \leq \frac{1}{2}n_g $, Fig. \ref{Fig_ConZono_Inapprox_Rand} shows the volume ratios for constrained zonotope, zonotope, and interval set inner-approximations. Both constrained zonotopes and zonotopes provide better approximations compared to interval sets while constrained zonotopes provide only a slightly higher mean volume ratio.
\end{exmp}

\section{Convex Hulls} \label{Sec_ConvexHull}
This section computes the CG-Rep of the convex hull of two constrained zonotopes $Z_1, Z_2 \subset \mathbb{R}^n$ with complexity $O(n+n_{c1}+n_{c2})$ where $n_{c1}$ and $n_{c2}$ are the number of constraints in $Z_1$ and $Z_2$, respectively. Since zonotopes are a subset of constrained zonotopes with $ n_c = 0 $, the following result also applies to zonotopes.

\begin{defn} \label{CH_def} \emph{\cite{Tiwary2008}} 
	The convex hull of the union of two polytopes $ P_1, P_2 \subset \mathbb{R}^n $ is defined as
	\begin{equation*}
	CH(P_1 \cup P_2) \triangleq \left\{\mathbf{x}_1 \lambda + \mathbf{x}_2 (1-\lambda) \mid \begin{matrix} \mathbf{x}_1 \in P_1,\\ \mathbf{x}_2 \in P_2, \\ 0 \leq \lambda \leq 1 \end{matrix}\right\}.
	\end{equation*}
\end{defn}
\begin{thm} \label{convexHull}
	The convex hull of the union of two constrained zonotopes $ Z_1 = \{\mathbf{G}_1,\mathbf{c}_1,\mathbf{A}_1,\mathbf{b}_1\} \subset \mathbb{R}^n $ and $ Z_2 = \{\mathbf{G}_2,\mathbf{c}_2,\mathbf{A}_2,\mathbf{b}_2\} \subset \mathbb{R}^n $ is a constrained zonotope $ Z_h = \{\mathbf{G}_h,\mathbf{c}_h,\mathbf{A}_h,\mathbf{b}_h\} $ where
	\begin{equation*}
	\begin{matrix*}[l]
	\mathbf{G}_h = \begin{bmatrix}
	\mathbf{G}_1 & \mathbf{G}_2 & \frac{\mathbf{c}_1-\mathbf{c}_2}{2} & \mathbf{0}
	\end{bmatrix}, & 
	
	\mathbf{c}_h = \frac{\mathbf{c}_1+\mathbf{c}_2}{2}, \\
	
	\mathbf{A}_h = \begin{bmatrix}
	\mathbf{A}_1 & \mathbf{0} & -\frac{\mathbf{b}_1}{2} & \mathbf{0} \\
	\mathbf{0} & \mathbf{A}_2 & \phantom{-}\frac{\mathbf{b}_2}{2} & \mathbf{0} \\
	\mathbf{A}_{3,1} & \mathbf{A}_{3,2} & \mathbf{A}_{3,0} & \mathbf{I}
	\end{bmatrix}, &
	
	\mathbf{b}_h = \begin{bmatrix} \phantom{-}\frac{1}{2} \mathbf{b}_1 \\ \phantom{-}\frac{1}{2} \mathbf{b}_2 \\ -\frac{1}{2} \mathbf{1}	\end{bmatrix}, \\
	
	\mathbf{A}_{3,1} = \begin{bmatrix} \phantom{-} \mathbf{I} \\ -\mathbf{I} \\ \phantom{-} \mathbf{0} \\ \phantom{-} \mathbf{0} \end{bmatrix},
	
	\mathbf{A}_{3,2} = \begin{bmatrix} \phantom{-} \mathbf{0} \\ \phantom{-}\mathbf{0} \\ \phantom{-}\mathbf{I} \\ -\mathbf{I} \end{bmatrix}, &
	
	\mathbf{A}_{3,0} = \begin{bmatrix} -\frac{1}{2}\mathbf{1} \\ -\frac{1}{2}\mathbf{1} \\ \phantom{-}\frac{1}{2}\mathbf{1} \\ \phantom{-}\frac{1}{2}\mathbf{1} \end{bmatrix}.
	\end{matrix*}
	\end{equation*}
\end{thm}
\begin{pf}
    Considering any element $\mathbf{x} \in Z_h$, it is to be proven that $\mathbf{x} \in CH(Z_1 \cup Z_2)$. By the definition of $Z_h$, $\exists \; \boldsymbol{\xi}_1 \in \mathbb{R}^{n_{g1}}, \boldsymbol{\xi}_2 \in \mathbb{R}^{n_{g2}}, \xi_0 \in \mathbb{R}$, and $\boldsymbol{\xi}_s \in \mathbb{R}^{2(n_{g1} + n_{g2})}$ such that
    \begin{subequations}\label{eq_Zh_exist_all}
        \begin{align}
            & \scriptsize{\mathbf{x} = \mathbf{G}_1\boldsymbol{\xi}_1+\mathbf{G}_2\boldsymbol{\xi}_2+\frac{\mathbf{c}_1-\mathbf{c}_2}{2}\xi_0+\mathbf{0}\boldsymbol{\xi}_s +\frac{\mathbf{c}_1+\mathbf{c}_2}{2},}  \label{eq_Zh_exist_element} \\
            & ||\boldsymbol{\xi}_1||_{\infty} \leq 1, \; ||\boldsymbol{\xi}_2||_{\infty} \leq 1, \; |\xi_0| \leq 1, \; ||\boldsymbol{\xi}_s||_{\infty} \leq 1, \label{eq_Zh_inf_norm_const} \\
            & \mathbf{A}_h [
            \boldsymbol{\xi}_1^T \; \boldsymbol{\xi}_2^T \; \xi_0 \; \boldsymbol{\xi}_s^T
            ]^T = \mathbf{b}_h. \label{eq_Zh_exist_cons}
        \end{align}
    \end{subequations}
    To prove $\mathbf{x} \in CH(Z_1 \cup Z_2)$ requires the existence of elements $\mathbf{z}_1$, $\mathbf{z}_2 \in \mathbb{R}^n$, $\lambda \in \mathbb{R}, \boldsymbol{\xi}_1^{'} \in \mathbb{R}^{n_{g1}}$, and $\boldsymbol{\xi}_2^{'} \in \mathbb{R}^{n_{g2}}$ such that
\begin{subequations}\label{eq_CH_Z1_Z2_exist_all}
    \begin{align}
        & \mathbf{x} = \mathbf{z}_1\lambda + \mathbf{z}_2(1 - \lambda), \quad 0 \leq \lambda \leq 1, \label{eq_CH_Z1_Z2_exp_base} \\ 
        & \mathbf{z}_1 = \mathbf{c}_1 + \mathbf{G}_1\boldsymbol{\xi}_1^{'}, \quad ||\boldsymbol{\xi}_1^{'}||_{\infty} \leq 1, \quad \mathbf{A}_1\boldsymbol{\xi}_1^{'} = \mathbf{b}_1, \label{eq_Zh_exist_element_Z1}\\
        & \mathbf{z}_2 = \mathbf{c}_2 + \mathbf{G}_2\boldsymbol{\xi}_2^{'}, \quad ||\boldsymbol{\xi}_2^{'}||_{\infty} \leq 1, \quad \mathbf{A}_2\boldsymbol{\xi}_2^{'} = \mathbf{b}_2. \label{eq_Zh_exist_element_Z2}
    \end{align}
\end{subequations}
This is shown by defining $\lambda$, $\boldsymbol{\xi}_1^{'}$, and $\boldsymbol{\xi}_2^{'}$ as
\begin{equation}\label{eq_cvxhull_lambda_xi_varsubs}
    \lambda = \frac{1}{2}(1 + \xi_0), \quad \boldsymbol{\xi}_1 = \boldsymbol{\xi}_1^{'}\lambda, \quad \boldsymbol{\xi}_2 = \boldsymbol{\xi}_2^{'} (1 - \lambda).
\end{equation}
By rearranging \eqref{eq_Zh_exist_element}, substituting using the variable definitions in \eqref{eq_cvxhull_lambda_xi_varsubs}, and then rearranging to simplify using the definitions for $\mathbf{z}_1$ and $\mathbf{z}_2$ from \eqref{eq_Zh_exist_element_Z1} and \eqref{eq_Zh_exist_element_Z2}, the expression for $\mathbf{x}$ from \eqref{eq_CH_Z1_Z2_exp_base} can be established as
\begin{subequations}
    \begin{align}
        \mathbf{x} &= \frac{\mathbf{c}_1}{2}(1 + \xi_0) + \mathbf{G}_1\boldsymbol{\xi}_1 + \frac{\mathbf{c}_2}{2}(1 - \xi_0) + \mathbf{G}_2\boldsymbol{\xi}_2, \label{eq_Zh_exist_element_exp1}\\ 
        &= \mathbf{c}_1\lambda + \mathbf{G}_1\boldsymbol{\xi}_1^{'}\lambda + \mathbf{c}_2(1 - \lambda) + \mathbf{G}_2\boldsymbol{\xi}_2^{'}(1 - \lambda), \label{eq_Zh_exist_element_exp2}\\ 
        &= \mathbf{z}_1\lambda + \mathbf{z}_2(1 - \lambda).
    \end{align}
\end{subequations}
Since $|\xi_0| \leq 1$, the definition for $\lambda$ in \eqref{eq_cvxhull_lambda_xi_varsubs} results in $0 \leq \lambda \leq 1$.
From the definition of $\mathbf{A}_h$ and $\mathbf{b}_h$, the first two sets of equality constraints are
\begin{equation}\label{eq_cvxhull_Zh_cons}
    \mathbf{A}_1\boldsymbol{\xi}_1 - \frac{\mathbf{b}_1}{2}\xi_0 = \frac{1}{2}\mathbf{b}_1, \quad \mathbf{A}_2\boldsymbol{\xi}_2 - \frac{\mathbf{b}_2}{2}\xi_0 = \frac{1}{2}\mathbf{b}_2.
\end{equation}
Using \eqref{eq_cvxhull_lambda_xi_varsubs}, \eqref{eq_cvxhull_Zh_cons} simplifies to
\begin{equation*}
    \mathbf{A}_1\boldsymbol{\xi}_1^{'} = \mathbf{b}_1, \; \forall \; \lambda \in (0, 1], \quad \mathbf{A}_2\boldsymbol{\xi}_2^{'} = \mathbf{b}_2, \; \forall \; \lambda \in [0, 1).
\end{equation*}
Note that if $\lambda = 0$, then $\mathbf{z}_1$ does not affect $\mathbf{x}$ or $\boldsymbol{\xi}_1$ and an arbitrary value of $\boldsymbol{\xi}_1^{'}$ can be chosen satisfying the infinity norm and equality constraints from \eqref{eq_Zh_exist_element_Z1}. Similarly, if $\lambda = 1$, an arbitrary value of $\boldsymbol{\xi}_2^{'}$ can be chosen satisfying constraints from \eqref{eq_Zh_exist_element_Z2}. Otherwise, the norm constraints $||\boldsymbol{\xi}_1^{'}||_{\infty} \leq 1$ and $||\boldsymbol{\xi}_2^{'}||_{\infty} \leq 1$ are guaranteed since $||\boldsymbol{\xi}_1||_{\infty} \leq 1$, $||\boldsymbol{\xi}_2||_{\infty} \leq 1$, and $0 \leq \lambda \leq 1$. Thus, $\mathbf{x} \in CH(Z_1 \cup Z_2)$.

Next, considering any $\mathbf{x} \in CH(Z_1 \cup Z_2)$, it is to be proven that $\mathbf{x} \in Z_h$.
By \textbf{Definition \ref{CH_def}}, there exists elements $\mathbf{z}_1$, $\mathbf{z}_2 \in \mathbb{R}^n$, $\lambda \in \mathbb{R}, \boldsymbol{\xi}_1^{'} \in \mathbb{R}^{n_{g1}}$, and $\boldsymbol{\xi}_2^{'} \in \mathbb{R}^{n_{g2}}$ such that \eqref{eq_CH_Z1_Z2_exp_base}-\eqref{eq_Zh_exist_element_Z2} hold.
To prove $\mathbf{x} \in Z_h$ requires the existence of variables $\boldsymbol{\xi}_1 \in \mathbb{R}^{n_{g1}}$, $\boldsymbol{\xi}_2 \in \mathbb{R}^{n_{g2}}$, $\xi_0 \in \mathbb{R}$, $\boldsymbol{\xi}_s \in \mathbb{R}^{2(n_{g1} + n_{g2})}$ such that \eqref{eq_Zh_exist_element}-\eqref{eq_Zh_exist_cons} hold.
Consider the following definitions for variables $\boldsymbol{\xi}_1$, $\boldsymbol{\xi}_2$, ${\xi}_0$, and $\boldsymbol{\xi}_s$ with
\begin{subequations}\label{eq_xi1_xi2_xi_0_xi_s_assum_all}
    \begin{align}
        & \boldsymbol{\xi}_1 = \boldsymbol{\xi}_1^{'}\lambda, \quad \boldsymbol{\xi}_2 = \boldsymbol{\xi}_2{'}(1-\lambda), \quad \xi_0 = 2\lambda -1, \label{eq_xi1_xi2_xi_0_assum}\\ 
        & \boldsymbol{\xi}_s = -\frac{1}{2}\mathbf{1} - (\mathbf{A}_{31}\boldsymbol{\xi}_1 + A_{32}\boldsymbol{\xi}_2 + \mathbf{A}_{30}\xi_0). \label{eq_xi_s_assum}
    \end{align}
\end{subequations} 
Using \eqref{eq_xi1_xi2_xi_0_assum} and \eqref{eq_xi_s_assum}, it can be readily shown that the equality constraints in \eqref{eq_CH_Z1_Z2_exp_base}-\eqref{eq_Zh_exist_element_Z2} can be rewritten to achieve \eqref{eq_Zh_exist_element} and \eqref{eq_Zh_exist_cons}. Thus, all that remains is to show $\big|\big| \; [
\boldsymbol{\xi}_1^T \; \boldsymbol{\xi}_2^T \; \xi_0 \; \boldsymbol{\xi}_s^T ]^T\; \big|\big|_{\infty} \leq 1$. Since $0 \leq \lambda \leq 1$ holds, 
\begin{equation*}
||\boldsymbol{\xi}_1^{'}||_{\infty} \geq ||\boldsymbol{\xi}_1^{'}\lambda||_{\infty} = ||\boldsymbol{\xi}_1||_{\infty},
\end{equation*}
is satisfied. By \eqref{eq_Zh_exist_element_Z1}, $||\boldsymbol{\xi}_1^{'}||_{\infty} \leq 1$ implies $||\boldsymbol{\xi}_1||_{\infty} \leq 1$. Similarly, it can be shown that $||\boldsymbol{\xi}_2||_{\infty} \leq 1$. Using the definition of $\xi_0$ from \eqref{eq_xi1_xi2_xi_0_assum} and $0 \leq \lambda \leq 1$ proves that $|\xi_0| \leq 1$. Finally, using the definition of $\boldsymbol{\xi}_s$ from \eqref{eq_xi_s_assum} and interval arithmetic, it can be shown that 
\begin{equation*}
||\boldsymbol{\xi}_1||_{\infty} \leq 1, \; ||\boldsymbol{\xi}_2||_{\infty} \leq 1, \; |\boldsymbol{\xi}_0|\leq 1 \implies ||\boldsymbol{\xi}_s||_{\infty} \leq 1.
\end{equation*} Thus, $\forall \; \mathbf{x} \in CH(Z_1 \cup Z_2)$, $\mathbf{x} \in Z_h$. \hfill \hfill \qed
\end{pf}

The resulting constrained zonotope $Z_h$ obtained using \textbf{Theorem \ref{convexHull}} has $n_{gh} = 3(n_{g1}+n_{g2})+1$ generators and $n_{ch} = n_{c1}+n_{c2}+2(n_{g1}+n_{g2})$ constraints.

\begin{figure}
	\begin{center}
		\includegraphics[width=8.6cm]{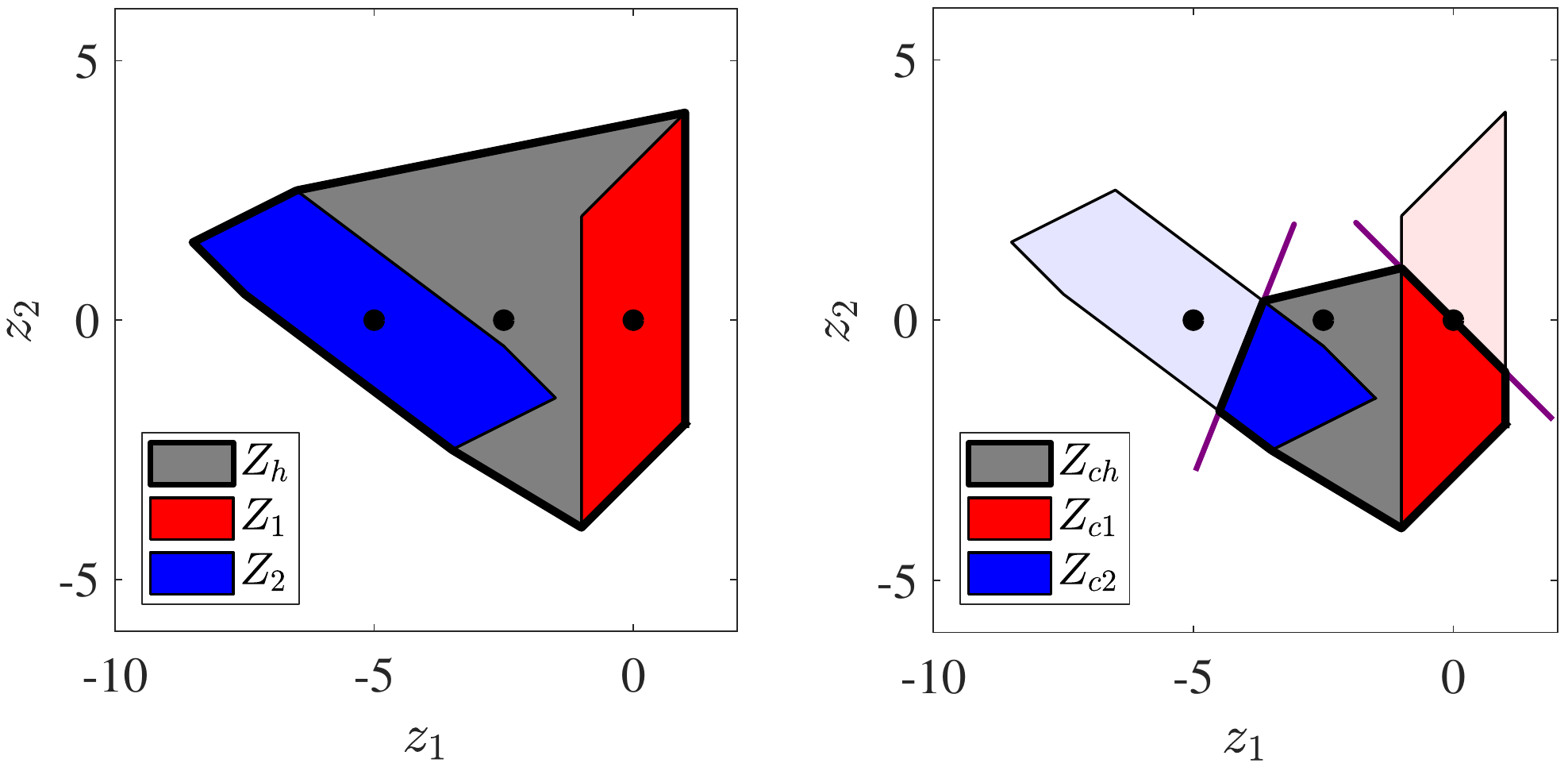}
		\caption{Left: The convex hull $ Z_h $ of zonotopes $ Z_1 $ and $ Z_2 $. 
			Right: The convex hull $ Z_{ch} $ of constrained zonotopes $ Z_{c1} $ and $ Z_{c2} $, where each constrained zonotope is a zonotope-halfspace intersection corresponding to the shown hyperplanes.} 
		\label{Fig_CVX_Hull}                      
	\end{center}                         
\end{figure}

\begin{exmp}
	For the zonotopes
	\begin{subequations} 
		\begin{align*}
		Z_1 &= \left\{ \begin{bmatrix} 0 & 1 & 0 \\ 1 & 1 & 2 \end{bmatrix}, \begin{bmatrix} 0 \\ 0	\end{bmatrix} \right\}, \\ 
		Z_2 &= \left\{ \begin{bmatrix} -0.5 & 1 & -2\\ \phantom{-}0.5 & 0.5 & 1.5\end{bmatrix}, \begin{bmatrix} -5 \\ \phantom{-}0	\end{bmatrix} \right\},
		\end{align*}
	\end{subequations}
	\setcounter{equation}{\value{equation}-1}
	Fig. \ref{Fig_CVX_Hull} shows the convex hull $ Z_h = CH(Z_1 \cup Z_2) $ with $n_g = 19 $ generators and $ n_c = 12 $ constraints, as computed using \textbf{\emph{Theorem \ref{convexHull}}}. Fig. \ref{Fig_CVX_Hull} also shows the convex hull $ Z_{ch} = CH(Z_{c1} \cup Z_{c2}) $ with $n_g = 25 $ generators and $ n_c =~18 $ constraints, where $ Z_{c1} = Z_1 \cap H_{1-} $, $ Z_{c2} = Z_2 \cap H_{2-} $, $ H_{1-} = \{\mathbf{z} \mid [1 \; 1] \mathbf{z} \leq 0 \} $, and $ H_{2-} = \{\mathbf{z}\mid [-2.5 \; 1] \mathbf{z} \leq 9.5 \} $.
\end{exmp}

\section{Robust Positively Invariant (RPI) Sets} \label{Sec_RPI}
This section provides both iterative and one-step optimization based methods for computing approximations of the minimal robust positively invariant set using zonotopes. Consider the autonomous discrete-time linear time-invariant system
\begin{equation} \label{autoSys}
\mathbf{x}_{k+1} = \mathbf{A} \mathbf{x}_k + \mathbf{w}_k,
\end{equation}
where $ \mathbf{x}_k \in \mathbb{R}^n $, $ \mathbf{A} \in \mathbb{R}^{n \times n} $ is a strictly stable matrix, and $ \mathbf{w}_k \in W \subset \mathbb{R}^n $, where $ W $ is a convex and compact set containing the origin.  

\begin{defn} \emph{\cite{Blanchini1999}}
	The set $ \Omega \subset \mathbb{R}^n $ is a robust positively invariant (RPI) set of \eqref{autoSys} if and only if $ \mathbf{A} \Omega \oplus W \subseteq \Omega $.
\end{defn}

\begin{defn} \emph{\cite{Rakovic2005}}
	The minimal RPI (mRPI) set $ F_\infty $ of \eqref{autoSys} is the RPI set that is contained in every closed RPI set of \eqref{autoSys} and is given by
	\begin{equation} \label{RPI_infsum}
	F_\infty = \bigoplus_{i=0}^{\infty} \mathbf{A}^i W.
	\end{equation}
\end{defn}

\subsection{Iterative Method}
Unless specific conditions are met, such as $ \mathbf{A} $ being nilpotent, the infinite sequence of Minkowski sums in \eqref{RPI_infsum} makes it impossible to compute $ F_\infty $ exactly. Thus, outer-approximations of the mRPI set are typically used. An iterative approach is developed in \cite{Rakovic2005} that computes the RPI set $ F(\alpha,s) $ such that $ F_\infty \subseteq F(\alpha,s) \subseteq F_\infty \oplus \epsilon B_\infty $, where $ \epsilon $ is a user defined bound on the error of the approximation with $ s \in \mathbb{N}_+ $, $ \alpha \in [0,1) $ such that $ \mathbf{A}^s W \subseteq \alpha W $. Starting at $ s = 0 $, the approach increments $ s $ until the approximation error is less than $ \epsilon $, at which point $ F_s $ is computed as 
\begin{equation} \label{RPI_finitesum}
F_s = \bigoplus_{i=0}^{s} \mathbf{A}^i W,
\end{equation}
and $ F(\alpha,s) = (1-\alpha)^{-1} F_s $. The iterative algorithm in \cite{Rakovic2005} requires use of multiple support functions at each iteration. When $ W $ is expressed in H-Rep, an LP must be solved for each support function calculation. As discussed in \cite{Trodden2016_OneStep}, computing $ F(\alpha,s) $ using this method may require the solution of thousands of LPs, even for a system with only two states. As briefly mentioned in Remark 3 in \cite{Rakovic2005}, if $ W $ is expressed in G-Rep, then the support function can be evaluated algebraically without the use of an LP, significantly reducing the computational cost. Thus, the use of zonotopes for RPI set calculations provides both improved scalability and reduced computational cost for the Minkowski sums in \eqref{RPI_finitesum} and by removing the need to solve LPs.

\subsection{One-step Optimization Method}

As an alternative for the iterative method in \cite{Rakovic2005}, a one-step method for computing an outer-approximation of the mRPI set is presented in \cite{Trodden2016_OneStep}. By expressing the RPI set in H-Rep, this method requires solving a single LP, assuming both the number and normal vectors of the hyperplanes associated with each halfspace inequality are provided \emph{a priori}. Inspired by this approach, the following presents a similar one-step method 
for computing an outer-approximation of the mRPI set using G-Rep, where the generator vectors are predetermined. 
\begin{thm} \label{RPI_OneStep}
	The zonotope $ Z = \{\mathbf{G}\mathbf{\Phi},\mathbf{c}\} \subset \mathbb{R}^n $, with $ \mathbf{\Phi} = diag(\boldsymbol{\phi}), \phi_i > 0, \forall i \in \{1, \cdots, n_g \} $, is an RPI set of \eqref{autoSys} if $ W = \{\mathbf{G}_w,\mathbf{c}_w\} $ and there exists $ \mathbf{\Gamma}_1 \in \mathbb{R}^{n_g \times n_g} $, $ \mathbf{\Gamma}_2 \in \mathbb{R}^{n_g \times n_w} $, and $ \boldsymbol{\beta} \in \mathbb{R}^{n_g} $ such that
	\begin{subequations} \label{RPI_conditions}
		\begin{align} 
		\mathbf{A}\mathbf{G}\mathbf{\Phi} &= \mathbf{G}\mathbf{\Gamma}_1, \label{RPI_condition1}\\
		\mathbf{G}_w &= \mathbf{G}\mathbf{\Gamma}_2, \label{RPI_condition2}\\
		(\mathbf{I} - \mathbf{A})\mathbf{c} - \mathbf{c}_w &= \mathbf{G}\boldsymbol{\beta}, \label{RPI_condition3} \\
		|\mathbf{\Gamma}_1|\mathbf{1} + |\mathbf{\Gamma}_2|\mathbf{1} + |\boldsymbol{\beta}| &\leq \mathbf{\Phi}\mathbf{1}. \label{RPI_condition4}
		\end{align}
	\end{subequations}
\end{thm}
\begin{pf}
	The proof requires showing that \eqref{RPI_conditions} enforces the zonotope containment conditions from \textbf{Lemma \ref{zonoContainment}} such that $ X \subseteq Y $, where $ X = \mathbf{A} Z \oplus W $ and $ Y = Z $. Consider the change of variables $ \mathbf{\Gamma}_1 = \mathbf{\Phi}\tilde{\mathbf{\Gamma}}_1 $, $ \mathbf{\Gamma}_2 = \mathbf{\Phi}\tilde{\mathbf{\Gamma}}_2 $, $ \boldsymbol{\beta} = \mathbf{\Phi}\tilde{\boldsymbol{\beta}} $ and define $ \tilde{\mathbf{\Gamma}} = [\tilde{\mathbf{\Gamma}}_1 \; \tilde{\mathbf{\Gamma}}_2] $. Then the zonotope containment conditions from \eqref{zonoContainment_Conditions} are satisfied by 1) rearranging and combining \eqref{RPI_condition1} and \eqref{RPI_condition2} to get $ [\mathbf{A}\mathbf{G}\mathbf{\Phi} \; \mathbf{G}_w] = \mathbf{G}\mathbf{\Phi} \tilde{\mathbf{\Gamma}} $, 2) rearranging \eqref{RPI_condition3} to get $ \mathbf{c} - (\mathbf{A}\mathbf{c} + \mathbf{c}_w) = \mathbf{G}\mathbf{\Phi}\tilde{\boldsymbol{\beta}} $, and 3) multiplying \eqref{RPI_condition4} by $ \mathbf{\Phi}^{-1} $, since $ \phi_i > 0 $, to get $ |\tilde{\mathbf{\Gamma}}|\mathbf{1} + |\tilde{\boldsymbol{\beta}}| \leq \mathbf{1} $. \hfill \hfill \qed
\end{pf}

\begin{figure}
	\begin{center}
		\includegraphics[width=8.6cm]{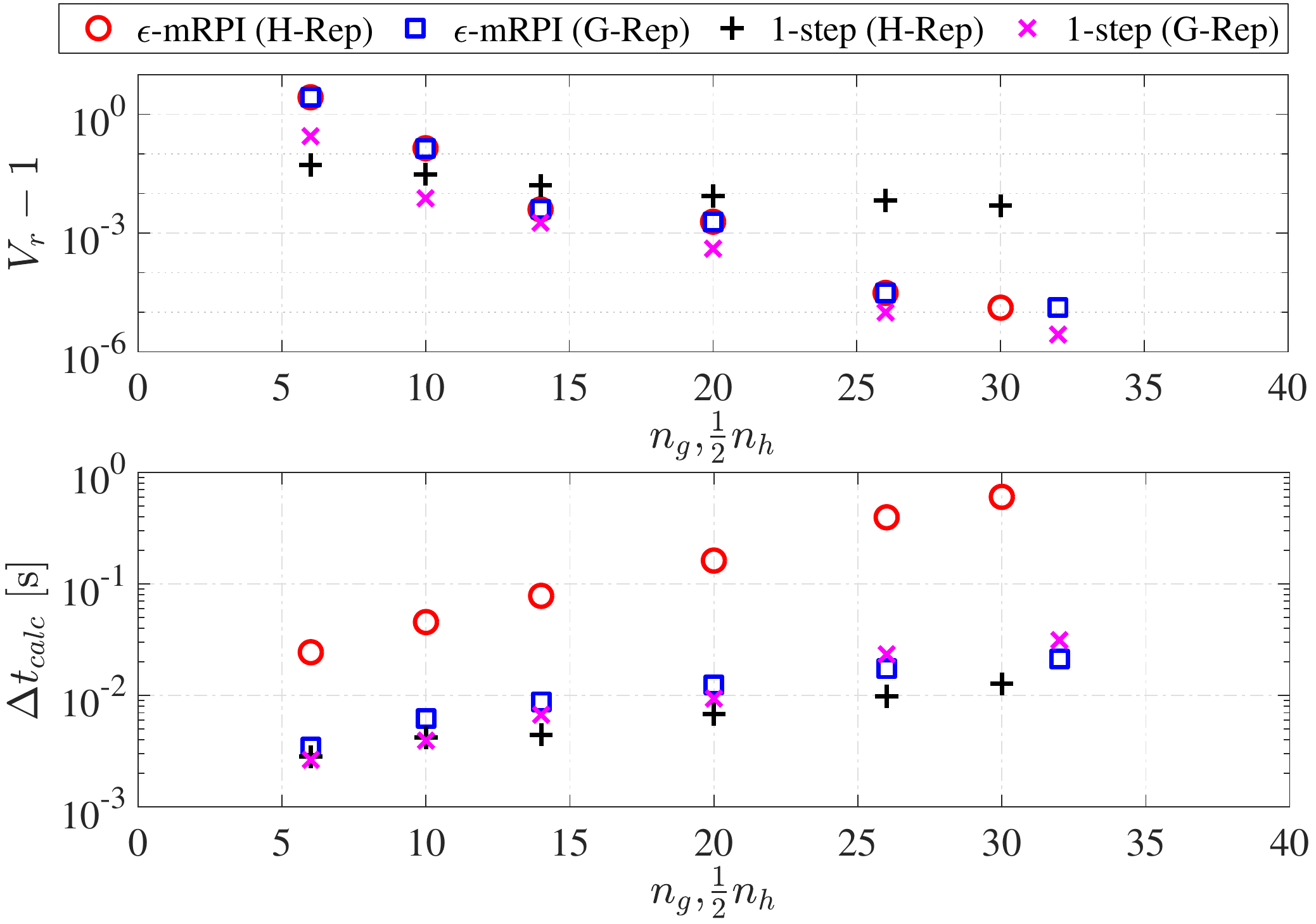}
		\caption{Comparison of volume ratio and computation time as a function of set complexity for outer-approximations of the mRPI set using iterative and 1-step approaches based on H-Rep or G-Rep.}
		\label{Fig_RPI_Set}                    
	\end{center}                         
\end{figure}

\begin{figure}
	\begin{center}
		\includegraphics[width=8.6cm]{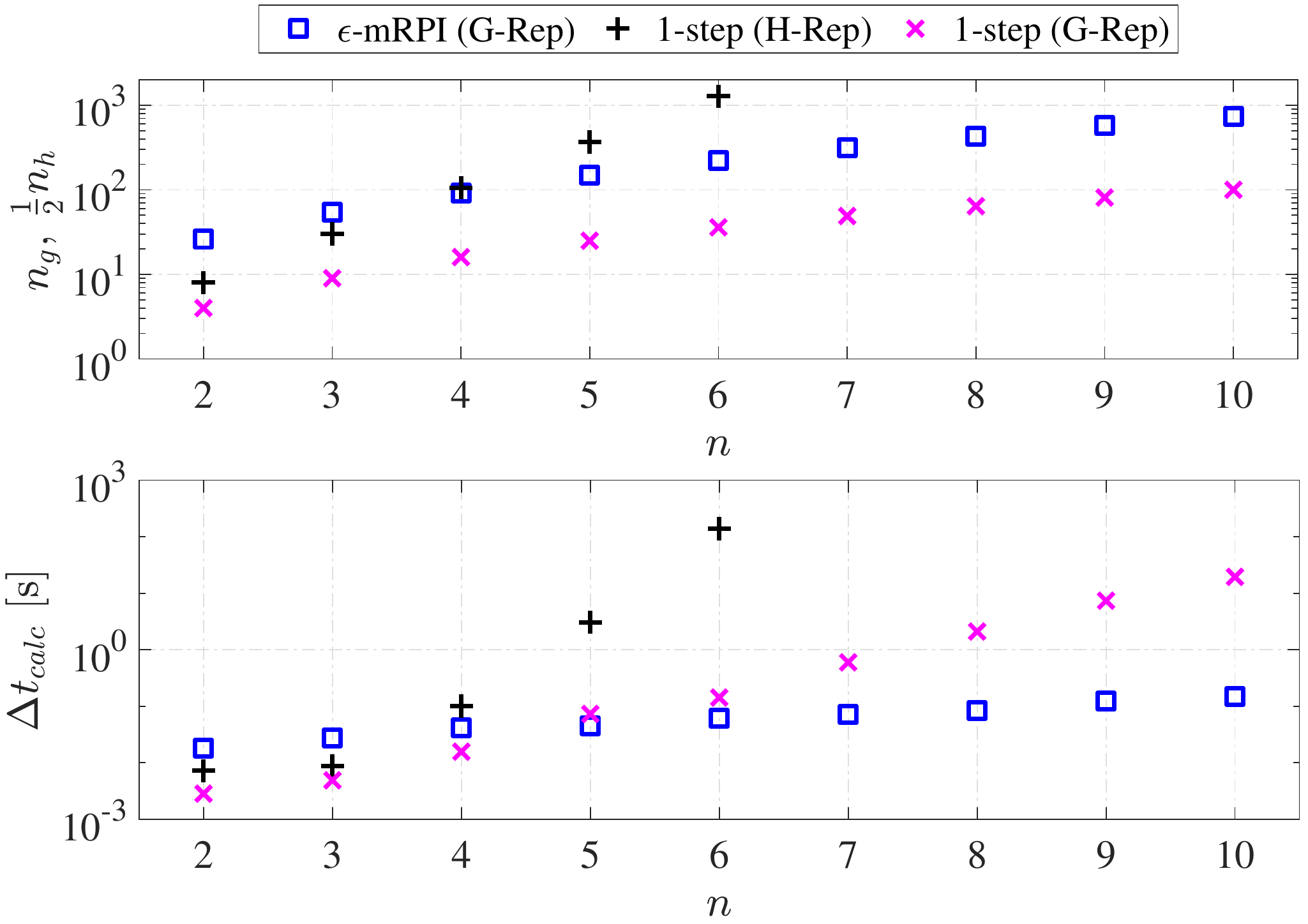}
		\caption{Comparison of set complexity and computation time as a function of system order for outer-approximations of the mRPI set using iterative and 1-step approaches based on H-Rep or G-Rep.}
	\label{Fig_RPI_Set_Scalability}         
	\end{center}
\end{figure}

When using \textbf{Theorem \ref{RPI_OneStep}} to determine the RPI set $ Z $ in G-Rep, the generator matrix $ \mathbf{G} $ is assumed to be known \emph{a~priori} in the same way that the normal vectors are chosen \emph{a~priori} in \cite{Trodden2016_OneStep} for the one-step RPI set computation in H-Rep. Given a desired order of $ Z $, $ \mathbf{G} $ can be computed using \eqref{RPI_finitesum} where $ \mathbf{G} = [\mathbf{G}_w \; \mathbf{A}\mathbf{G}_w \; ... \; \mathbf{A}^s\mathbf{G}_w] $, for some $ s \in \mathbb{N}_+ $ that provides the desired order. 
Once $ \mathbf{G} $ is determined, the diagonal matrix $ \mathbf{\Phi} $ provides the ability to scale the size of $ Z $ such that $ Z $ is an RPI set. Since the minimal RPI set is typically desired, an optimization problem can be formulated with the constraints from \eqref{RPI_conditions} and a objective function that minimizes the scaling variables in $ \mathbf{\Phi} $. With $ \mathbf{c} $, $ \mathbf{\Phi} $, $ \mathbf{\Gamma}_1 $, $ \mathbf{\Gamma}_2 $, and $ \boldsymbol{\beta} $ as decision variables in this optimization problem, \eqref{RPI_conditions} consists of only linear constraints and thus an LP or QP can be formulated based on the norm used to minimize the vector $ \boldsymbol{\phi} $, where $ \mathbf{\Phi} = diag(\boldsymbol{\phi}) $. In the following example, an LP is formulated by minimizing $ \| \boldsymbol{\phi} \|_\infty $ subject to \eqref{RPI_conditions}. Computing RPI set $Z$ using \textbf{Theorem \ref{RPI_OneStep}} requires solving an LP with $n_g^2 + n_g(n_w + 2) + n$ decision variables.
\begin{exmp}
	Consider the system from \emph{\cite{Trodden2016_OneStep}}
	\begin{equation}\label{RPISet_Trodden_Sys}
	\mathbf{x}_{k+1} = \begin{bmatrix}	1 & 1 \\ 0 & 1	\end{bmatrix}\mathbf{x}_k + \begin{bmatrix} 0.5 \\ 1
	\end{bmatrix} u_k + \mathbf{w}_k, 
	\end{equation} 
	with $ \mathbf{w}_k \in W = \{ \mathbf{w} \in \mathbb{R}^2 \mid \| \mathbf{w} \|_{\infty} \leq 0.1\} $. As in \emph{\cite{Trodden2016_OneStep}}, the state feedback control law $ u_k = \mathbf{K} \mathbf{x}_k $, where $\mathbf{K} $ corresponds to the LQR solution with $ \mathbf{Q} = \mathbf{I} $ and $ \mathbf{R} = 1 $, converts \eqref{RPISet_Trodden_Sys} to an autonomous system of the form \eqref{autoSys}. For this  system, four methods for computing outer-approximations of the mRPI set are compared in Fig. \ref{Fig_RPI_Set} with respect to volume ratio $ V_r $ and computation time $ \Delta t_{calc} $ as a function of set complexity ($ n_g $~for zonotopes in G-Rep, $ \frac{1}{2}n_h $ for polytopes in H-Rep). The seminal work from \emph{\cite{Rakovic2005}}, denoted as $\epsilon$-mRPI (H-Rep), is the most computationally expensive since evaluating support functions for polytopes in H-Rep requires the solution of an LP. Using zonotopes in G-Rep, computational cost of this $\epsilon$-mRPI approach can be reduced by an order-of-magnitude since evaluating support functions for zonotopes is algebraic, as mentioned in Remark 3 of \emph{\cite{Rakovic2005}}. Alternatively, the 1-step approaches from \emph{\cite{Trodden2016_OneStep}} and \emph{\textbf{Theorem \ref{RPI_OneStep}}}, provide similar computational advantages. However, the 1-step approach from \emph{\cite{Trodden2016_OneStep}} is sensitive to the choice of hyperplanes. Using the same choice of hyperplanes from \emph{\cite{Trodden2016_OneStep}}, Fig. \ref{Fig_RPI_Set} shows that the volume ratio does not decrease with increasing set complexity as quickly as the zonotope-based approach. Note that volume ratio is defined with respect to an approximation of the true mRPI set volume computed using the $\epsilon$-mRPI method with $\epsilon = 10^{-9}$. 
	
	To assess the scalability of these methods with respect to system order, Fig. \ref{Fig_RPI_Set_Scalability} shows a comparison of these methods based on set complexity and computation time as a function of system order $ n $. Note that the  $\epsilon$-mRPI (H-Rep) method became impractical for higher system orders and is not included in Fig. \ref{Fig_RPI_Set_Scalability}. Similarly, the 1-step (H-Rep) method became impractical for $ n > 6 $. These results are generated using a $n^{th}$-order integrator system similar to that of \eqref{RPISet_Trodden_Sys}. While the $\epsilon$-mRPI method in G-Rep provides the lowest computational cost, the complexity of the resulting set is roughly ten times larger than the set used for the 1-step approach. While scaling better than the 1-step H-Rep approach, the 1-step G-Rep approach requires solving a linear program with the constraints from \eqref{RPI_conditions} which includes the large decision variable $ \mathbf{\Gamma}_1 \in \mathbb{R}^{n_g \times n_g} $. To manage this computational cost for higher order systems, the number of steps $ s \in \mathbb{N}_+ $ in \eqref{RPI_finitesum} can be chosen to balance set complexity and accuracy.
\end{exmp}


\section{Pontryagin Difference} \label{Sec_Pontryagin}

This section provides an iterative method for computing the constrained zonotope representation of the Pontryagin difference of two zonotopes and a one-step optimization method for computing the zonotopic inner-approximation of the Pontryagin difference.

\begin{defn} \emph{\cite{Althoff2015a}}
	Given two sets $ Z_1, Z_2 \subset \mathbb{R}^n $, the Pontryagin difference $ Z_d = Z_1 \ominus Z_2 $ is defined as
	\begin{equation} \label{Pontryagin_Diff_Def}
	Z_d = \{z \in \mathbf{R}^n \mid z \oplus Z_2 \subseteq Z_1\}.
	\end{equation}
\end{defn}
The Pontryagin difference is also referred to as the Minkowski difference or the erosion of set $ Z_1 $ by $ Z_2 $.

\subsection{Iterative Method}
If $ Z_1 $ and $ Z_2 $ are zonotopes, then \cite{Althoff2015a} provides the following iterative method for computing $Z_d$. 

\begin{lem} (Theorem 1 of  \emph{\cite{Althoff2015a}}) \label{Pontryagin_Diff}
	If $ Z_1 = \{\mathbf{G}_1,\mathbf{c}_1\} $ and $ Z_2 = \{\mathbf{G}_2,\mathbf{c}_2\} $, then the Pontryagin difference $ Z_d = Z_1 \ominus Z_2 $ is computed using the $ n_{g2} $ generators $ \mathbf{g}_{2,i} $ of $ Z_2 $ by applying the following recursion:
	\begin{subequations} \label{Pontryagin_Diff_calc}
		\begin{align}
		Z_{int}^{(0)} &= Z_1 - c_2, \\
		Z_{int}^{(i)} &= (Z_{int}^{(i-1)} + \mathbf{g}_{2,i}) \cap (Z_{int}^{(i-1)} - \mathbf{g}_{2,i}), \label{Pontryagin_intersection}\\
		Z_d &= Z_{int}^{(n_{g2})}.
		\end{align}
	\end{subequations}
\end{lem}

As shown in \cite{Althoff2015a}, zonotopes are not closed under the Pontryagin difference. Thus, the methods in \cite{Althoff2015a} require the use of a combination of G-Rep and H-Rep to compute approximations of $ Z_d $ in G-Rep. While this combination results in faster calculations than methods that solely use H-Rep, the majority of computation time comes from the conversion from G-Rep to H-Rep, which scales exponentially with the number of generators. 

However, since $ Z_d $ is computed via the intersection of zonotopes, $ Z_d $ can be exactly represented as a constrained zonotope. Thus, \eqref{Pontryagin_intersection} can be directly computed using the generalized intersection from \eqref{generalized_Intersection} without the need for H-Rep. Note that iterative method from \textbf{Lemma \ref{Pontryagin_Diff}} is also applicable if $ Z_1 = \{\mathbf{G}_1,\mathbf{c}_1,\mathbf{A}_1,\mathbf{b}_1\} $ is a constrained zonotope, since \eqref{Pontryagin_Diff_calc} only requires $ Z_2 $ to be the Minkowski sum of generators $ \mathbf{g}_{2,i} $. For a constrained zonotope $Z_1$ in $\mathbb{R}^n$ with $n_{c1}$ constraints and $n_{g1}$ generators and a zonotope $Z_2$ in $\mathbb{R}^n$ with $n_{g2}$ generators, $Z_d = Z_1 \ominus Z_2$ is a constrained zonotope with $n_{gd} = 2^{n_{g2}}n_{g1}$ generators and $n_{cd} = 2^{n_{g2}}n_{c1} + n(2^{n_{g2}} - 1)$ constraints.


\subsection{One-step Optimization Inner-Approximation Method}

As an alternative to the iterative method from \textbf{Lemma~\ref{Pontryagin_Diff}}, the following theorem presents a one-step method for computing an zonotopic inner-approximation of the Pontryagin difference $ \tilde{Z}_d \subseteq Z_d = Z_1 \ominus Z_2$ using a single LP.

\begin{thm} \label{Pontryagin_OneStep}
	Given $ Z_1 = \{\mathbf{G}_1,\mathbf{c}_1\} $ and $ Z_2 = \{\mathbf{G}_2,\mathbf{c}_2\} $, then $ \tilde{Z}_d = \{[\mathbf{G}_1 \; \mathbf{G}_2]\mathbf{\Phi},\mathbf{c}_d\} $, with $ \mathbf{\Phi} = diag(\boldsymbol{\phi}), \phi_i > 0, \forall i \in \{1, \cdots, n_{g1} + n_{g2}\}$, is an inner-approximation of the Pontryagin difference such that $ \tilde{Z}_d \subseteq Z_1 \ominus Z_2 $ if there exists $ \mathbf{\Gamma} \in \mathbb{R}^{n_{g1} \times (n_{g1} + 2 n_{g2})} $ and $ \boldsymbol{\beta} \in \mathbb{R}^{n_{g1}} $, such that
	\begin{subequations} \label{Pontryagin_conditions}
		\begin{align} 
		\begin{bmatrix}
		[\mathbf{G}_1 \; \mathbf{G}_2]\mathbf{\Phi} & \mathbf{G}_2
		\end{bmatrix} &= \mathbf{G}_1 \mathbf{\Gamma}, \label{Pontryagin_condition1}\\
		\mathbf{c}_1 - (\mathbf{c}_d + \mathbf{c}_2) &= \mathbf{G}_1 \boldsymbol{\beta}, \label{Pontryagin_condition2}\\
		|\mathbf{\Gamma}|\mathbf{1} + |\boldsymbol{\beta}| &\leq \mathbf{1}. \label{Pontryagin_condition3}
		\end{align}
	\end{subequations}
\end{thm}
\begin{pf}
	By viewing \eqref{Pontryagin_conditions} in the context of the zonotope containment conditions from \textbf{Lemma \ref{zonoContainment}}, it is clear that \eqref{Pontryagin_conditions} enforces the  Pontryagin difference condition $ \tilde{Z}_d \oplus Z_2 \subset Z_1 $ from \eqref{Pontryagin_Diff_Def}. \hfill \hfill \qed
\end{pf}

When using \textbf{Theorem \ref{Pontryagin_OneStep}} to compute $ \tilde{Z}_d \subset Z_d $ in G-Rep, the generator matrix $ [\mathbf{G}_1 \; \mathbf{G}_2]\mathbf{\Phi} $ is assumed to be comprised of the generators from both $ Z_1 $ and $ Z_2 $ scaled by the diagonal matrix $ \mathbf{\Phi} $. Since maximizing the size of $ \tilde{Z}_d $ is typically desired, an optimization problem can be formulated with the constraints from \eqref{Pontryagin_conditions} and an objective function that maximizes the scaling variables in $ \mathbf{\Phi} $. With $ \mathbf{c}_d $, $ \mathbf{\Phi} $, $ \mathbf{\Gamma} $, and $ \boldsymbol{\beta} $ as decision variables in this optimization problem, \eqref{Pontryagin_conditions} consists of only linear constraints and thus an LP or QP can be formulated based on the norm used to maximize the vector $ \boldsymbol{\phi} $, where $ \mathbf{\Phi} = diag(\boldsymbol{\phi})$. Computing $\tilde{Z}_d$ using \textbf{Theorem \ref{Pontryagin_OneStep}} requires solving a LP with $n_{g1}^2 +2n_{g1}n_{g2} + 2n_{g1} + n_{g2} + n$ decision variables.

\begin{figure}
	\begin{center}
		\includegraphics[width=7cm]{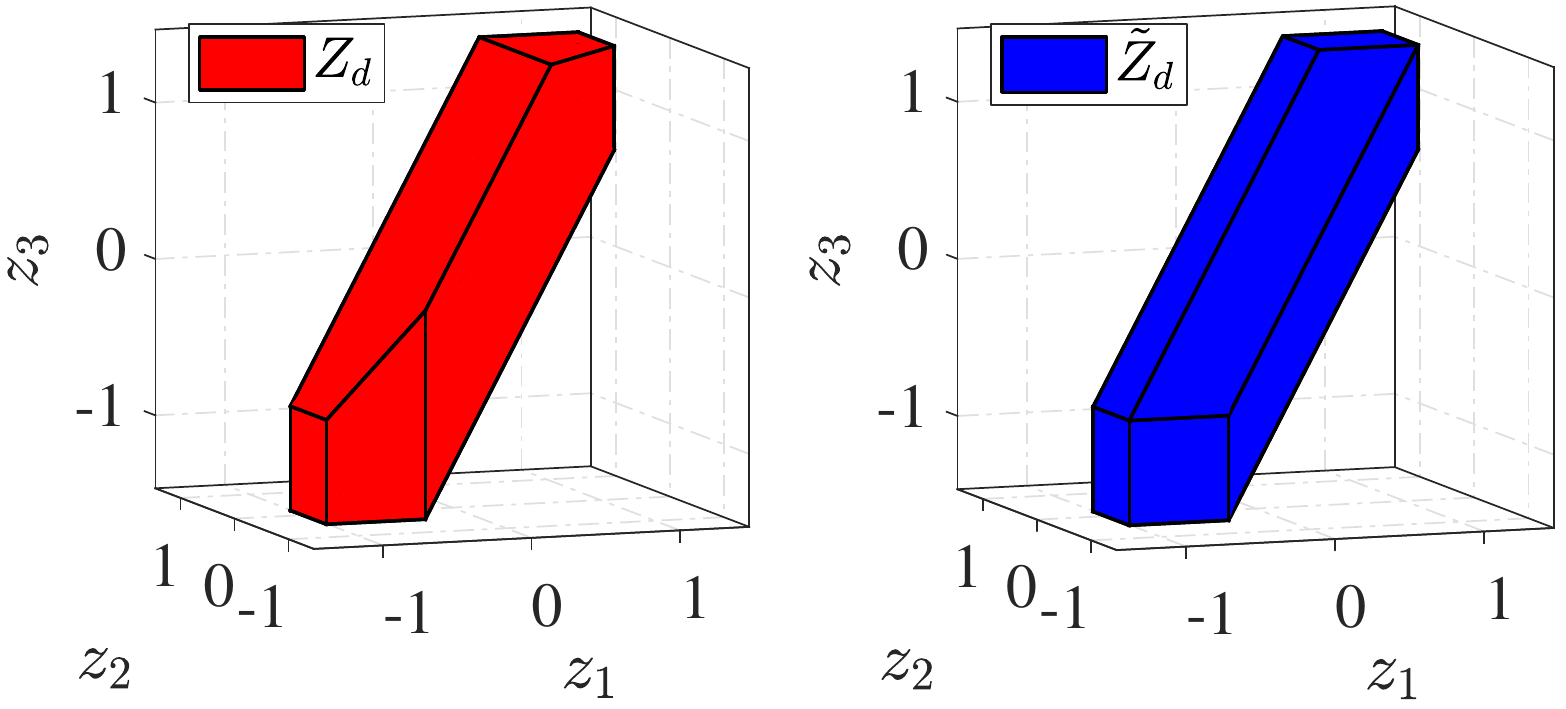}
		\caption{Left: The Pontryagin difference $ Z_d = Z_1 \ominus Z_2 $ where $ Z_1 $ and $ Z_2 $ are zonotopes but $ Z_d $ is not \cite{Althoff2015a}. Right: The inner-approximation of $ Z_d $ by a zonotope $ \tilde{Z}_d \subseteq Z_d $. }
		\label{Fig_Pontryagin_Diff_3D}                   
	\end{center}                     
\end{figure}

\begin{exmp}
	Consider the zonotopes from \emph{\cite{Althoff2015a}}
	\begin{equation*}
	\setlength\arraycolsep{2pt}
	Z_1 = \left\{ \begin{bmatrix} 1 & 1 & 0 & 0 \\ 1 & 0 & 1 & 0 \\ 1 & 0 & 0 & 1 \end{bmatrix}, \boldsymbol{0} \right\}, \quad Z_2 = \left\{ \frac{1}{3} \begin{bmatrix} -1 & 1 & 0 & 0 \\ \phantom{-}1 & 0 & 1 & 0 \\ \phantom{-}1 & 0 & 0 & 1 \end{bmatrix}, \boldsymbol{0} \right\}.
	\end{equation*}
	Fig. \ref{Fig_Pontryagin_Diff_3D} shows the Pontryagin difference $ Z_d = Z_1 \ominus Z_2 $ with $ n_g = 64 $ and $ n_c = 45 $ computed using \emph{\textbf{Lemma~\ref{Pontryagin_Diff}}}. As discussed in \emph{\cite{Althoff2015a}}, zonotopes are not closed under the Pontryagin difference, which can be seen in Fig. \ref{Fig_Pontryagin_Diff_3D} by the asymmetric facets of $ Z_d $. Using \emph{\textbf{Theorem \ref{Pontryagin_OneStep}}}, the inner-approximation of the Pontryagin difference $ \tilde{Z}_d $ is also shown in Fig. \ref{Fig_Pontryagin_Diff_3D}. Choosing to maximize $ \| [\mathbf{G}_1 \; \mathbf{G}_2]\mathbf{\Phi} \|_\infty $ subject to \eqref{Pontryagin_conditions} produced $ \tilde{Z}_d \subset Z_d $ with a volume ratio of $ V_r = 0.924 $.
\end{exmp}

\begin{table*}[!t]
	\renewcommand{\arraystretch}{1.1}
	\caption{Pontryagin difference set complexity and computation time (seconds)}
	\label{Tab_Ptryagindiff}
	\centering
	\begin{tabular}{ M{0.3cm} M{0.4cm}  M{0.4cm}|  M{0.6cm}  M{0.6cm} | M{3.5cm} M{0.9cm} | M{0.6cm} M{0.6cm} M{0.6cm} }
		\hline
		\multicolumn{3}{c}{} & \multicolumn{7}{c}{$Z_1 \ominus Z_2$} \\
		\cline{4-10} 
		\multicolumn{1}{c}{} & \multicolumn{1}{c}{$Z_1$} & \multicolumn{1}{c}{$Z_2$} & \multicolumn{2}{c}{H-Rep} & \multicolumn{2}{c}{CG-Rep} & \multicolumn{3}{c}{1-Step (G-Rep)} \\ \cline{2-3}  \cline{4-10}
		\multicolumn{1}{c}{$n$} & \multicolumn{2}{c}{$n_g$} & \multicolumn{1}{c}{$n_h$} & \multicolumn{1}{c}{$t_h$} & \multicolumn{1}{c}{$n_{c} \times n_g$} & \multicolumn{1}{c}{$t_h/t_{cg}$} & \multicolumn{1}{c}{$n_g$} & \multicolumn{1}{c}{$V_r$} & \multicolumn{1}{c}{$t_h/t_{g}$} \\
		\hline
		2 & 4  & 4  & 16    & 0.01 & 30 	 $\times$ 64  		& 33.2 	& 2.5 & 0.64 & 3.3\\
		2 & 8  & 4  & 32    & 0.01 & 30 	 $\times$ 128 		& 48.2 	& 3.0 & 0.54 & 3.5\\
		2 & 4  & 8  & 16    & 0.01 & 510	 $\times$ 1024 		& 17.6 	& 2.6 & 0.67 & 3.3\\
		2 & 8  & 8  & 32    & 0.02 & 510 	 $\times$ 2048  	& 20.9 	& 3.2 & 0.53 & 3.1\\ \hline 
		3 & 6  & 6  & 60    & 0.03 & 189  	 $\times$ 384 		& 68.8 	& 3.7 & 0.55 & 7.5 \\
		3 & 12 & 6  & 264   & 0.16 & 189 	 $\times$ 768 		& 261 	& 4.8 & 0.46 & 16.0 \\
		3 & 6  & 12 & 60    & 0.13 & 12,285  $\times$ 24,576 	& 18.6 	& 3.8 & 0.54 & 21.5\\
		3 & 12 & 12 & 264   & 0.40 & 12,285  $\times$ 49,152 	& 27.2	& 4.9 & 0.43 & 28.2\\ \hline 
		4 & 8  & 8  & 224   & 0.38 & 1,020   $\times$ 2,048  	& 359 	& 4.9 & 0.50 & 46.2 \\ 
		4 & 16 & 8  & 2,240 & 41.0 & 1,020   $\times$ 4,096  	& 2,370 & 6.2 & 0.45 & 1,890\\ 
		4 & 8  & 16 & 224   & 59.0 & 262,140 $\times$ 524,288   & 271 	& 4.7 & 0.43 & 4,078\\ 
		4 & 16 & 16 & 2,240 & 243  & 262,140 $\times$ 1,048,576 & 556 	& 6.2 & 0.48 & 6,510\\ \hline 
	\end{tabular}
\end{table*}

\begin{exmp}
	Similar to \emph{\cite{Althoff2015a}}, the scalability of exact constrained zonotope representations of the Pontryagin difference via \emph{\textbf{Lemma~\ref{Pontryagin_Diff}}} and  zonotopic inner-approximations via \emph{\textbf{Theorem~\ref{Pontryagin_OneStep}}} is compared with the standard H-Rep approach provided in the Multi-Parametric Toolbox \cite{Herceg2013}. Table \ref{Tab_Ptryagindiff} shows the complexity and computational time for computing the Pontryagin difference $ Z_d = Z_1 \ominus Z_2 $ using each of the three methods for zonotopes in $ \mathbb{R}^2 $, $ \mathbb{R}^3 $, and $ \mathbb{R}^4 $. Each entry in Table~\ref{Tab_Ptryagindiff} represents an average of 100 computations using randomly generated zonotopes $ Z_1 $ and $ Z_2 $. These random zonotopes are generated using the procedure provided in \emph{\cite{Althoff2015a}} and the CORA toolbox \cite{Althoff2018a}. Cases where $ Z_d = \emptyset $ were disregarded and not considered in the set of 100 computations. For CG-Rep and G-Rep, the ratio of computation times relative to that of H-Rep is presented. Since the G-Rep approach is an inner-approximation, the average volume ratio is also provided. From these results, it is clear that both the set complexity $ n_h $ and the computation time $ t_h $ for the H-Rep approach increase by approximately an order-of-magnitude as the set dimension $ n $ increases. While the CG-Rep approach increases the computation speed by approximately two orders-of-magnitude, the set complexity increases exponentially. Sparse matrices were used to reduce the memory requirements for these computations.  The redundancy removal approach presented in Section \ref{Sec_Redundancy} was not able to detect the high-degree of redundancy in these set representations.  Alternatively, the one-step G-Rep approximation approach also provided significant reductions in computational cost while maintaining a small number of generators. However, for these randomly generated zonotopes, the inner-approximation only captures approximately 50\% of the volume of $ Z_d $.  While these methods will likely work well for many practical applications, future work is needed to improve redundancy detection and removal for the CG-Rep approach and improved optimization formulations are needed for the G-Rep approach to further maximize volume ratio.
\end{exmp}
\section{Application to Reachability Analysis} \label{Sec_Hier}

To demonstrate the applicability of algorithms developed in this paper, this section considers the exact and approximate computations of backwards reachable sets of a constrained linear system in the context of the two-level hierarchical MPC framework developed in \cite{Koeln2019ACC,Koeln2019_Aut}. The high-level goal is to compute a \emph{wayset} $ Z_c(k) $ at discrete time step $ k $ that captures all of the initial states $ \mathbf{x}(k) \in Z_c(k) \subset \mathbb{R}^n $ for which there are state and input trajectories $ \mathbf{x}(k+j) $ and $ \mathbf{u}(k+j) $ that satisfy, for all $ j \in \{0,\cdots,N-1\} $, \emph{i}) the dynamics $ \mathbf{x}(k+j+1) = \mathbf{A} \mathbf{x}(k+j) + \mathbf{B} \mathbf{u}(k+j) $, \emph{ii}) the state and input constraints $ \mathbf{x}(k+j) \in \mathcal{X}$ and $ \mathbf{u}(k+j) \in \mathcal{U}$, and \emph{iii}) the terminal constraint $ \mathbf{x}(k+N) = \mathbf{x}^* $ for some predetermined target $ \mathbf{x}^* \in \mathbb{R}^n $. In the context of the hierarchical MPC framework from \cite{Koeln2019ACC,Koeln2019_Aut}, $ \mathbf{x}^* $ is a future state on the optimal trajectory determined by an upper-level controller and $ Z_c(k) $ is a terminal constraint imposed on a lower-level controller. Since $ \mathbf{x}^* $ is updated at every evaluation of the upper-level controller, $ Z_c(k) $ must be recomputed in real-time, which is enabled through the use of constrained zonotopes.

\begin{figure}[t]
	\begin{center}		\includegraphics[width=7cm]{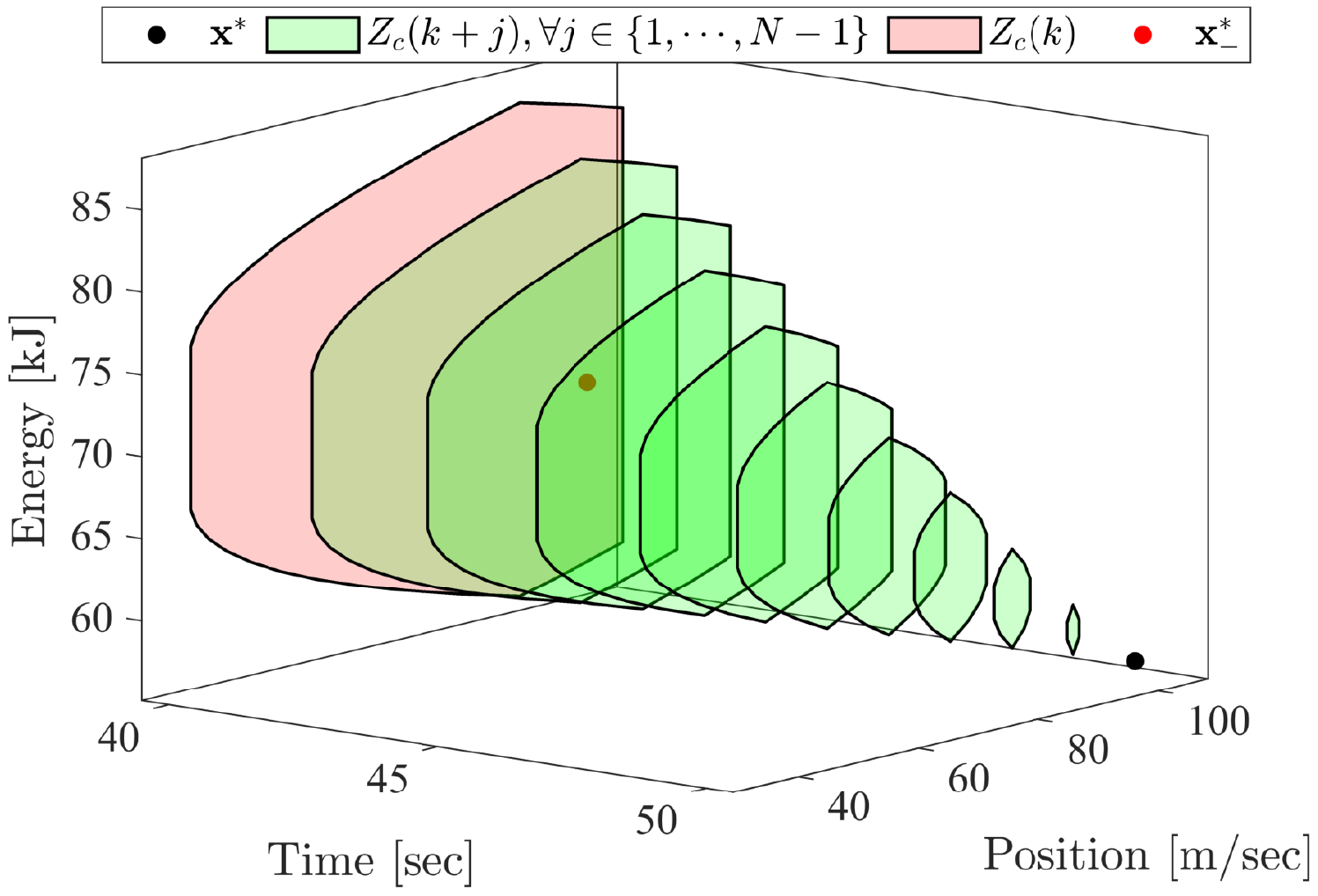}
	\caption{The evolution of backward reachable wayset $Z_c(k)$ for $k = 40$ and $N = 10$ time steps starting from $\mathbf{x}^*$ projected on the position and energy states. The sets $Z_c(k + j), \forall \; j \in \{ 7, 8, 9\}$ are zonotopes (evident from symmetry) while the sets $Z_c(k+j), \forall \; j \in \{ 0, \cdots, 6\}$, are constrained zonotopes. The constrained zonotope wayset $Z_c(k)$ contains $\mathbf{x}_{-}^*$ ensuring the control feasibility from \cite{Koeln2019ACC,Koeln2019_Aut}.}
	\label{Fig_HierMPC_Wayset_evol}   
	\end{center}        
\end{figure}

\textbf{Algorithm \ref{waysetComputation}} shows a simplified version of the backward reachable wayset algorithms presented in \cite{Koeln2019ACC,Koeln2019_Aut}. Fig. \ref{Fig_HierMPC_Wayset_evol} shows the results of this algorithm when applied to the simplified vehicle system model from \cite{Koeln2019ACC,Koeln2019_Aut} with
\begin{equation}
    \mathbf{x}(k+1) = \begin{bmatrix} 1 & 1 & 0 \\ 0 & 1 & 0 \\ 0 & 0 & 1 \end{bmatrix}\mathbf{x}(k) + \begin{bmatrix} \phantom{-}0 & \phantom{-}0 & \phantom{-}0 \\ \phantom{-}1 & -1 & \phantom{-}0 \\ -1 & -1 & -1
    \end{bmatrix}\mathbf{u}(k),
\end{equation}
where the states represent position, velocity, and on-board energy storage and the inputs represent acceleration, deceleration, and power to an on-board load. The  discretization time step size is $\Delta t = 1$ second and the state and input constraints defining $\mathcal{X}$ and $\mathcal{U}$ are
\begin{equation*}
    \begin{bmatrix} -1 \\ -20 \\ 0 \end{bmatrix} \leq \mathbf{x}(k) \leq \begin{bmatrix}
    105 \\ 20 \\ 100
    \end{bmatrix}, \; \begin{bmatrix} 0 \\ 0 \\ 0 \end{bmatrix} \leq \mathbf{u}(k) \leq \begin{bmatrix} 1 \\ 1 \\ 1\end{bmatrix}.
\end{equation*}

To demonstrate the halfspace intersection results from Section \ref{halfspaceIntersections}, Table \ref{Tab_wayset_cmplx_comptime} compares the set representation complexity and computation time of four different CG-Rep methods with those using H-Rep via the Multi-Parametric Toolbox \cite{Herceg2013}. All computation times are averaged over 100 runs. Overall, the CG-Rep methods result in significantly less set complexity and computation time. The CG-Rep methods differ in the computation of $ \hat{Z}_c(k + j-1) \cap \mathcal{X}$ in \textbf{Algorithm \ref{waysetComputation}}. Specifically, this intersection is computed using $1)$ the zonotope-hyperplane (ZH) method from \textbf{Lemma \ref{Zono_halfspace_int_check}} based on the parent zonotope $ \hat{Z}(k + j-1) \supset \hat{Z}_c(k + j-1) $ and the H-Rep of $ \mathcal{X} $, $2)$ the generalized intersection (GI) (from \eqref{generalized_Intersection}) of the constrained zonotope wayset and the G-Rep of $ \mathcal{X} $, $3)$ the linear program (LP) method from \textbf{Lemma \ref{conzono_hyplane_intersection}} for checking the intersection of a constrained zonotope and a hyperplane, and $4)$ the interval arithmetic (IA) approach using \textbf{Algorithm \ref{gen_Bounds}} to detect empty sets when $ \hat{Z}_c(k + j-1) \subset \mathcal{X}$.
In the ZH, LP, and IA methods, if the wayset intersects the hyperplanes associated with the halfspaces of $ \mathcal{X} $, generators and constraints are added using \eqref{conzono_halfspace_int} to exactly compute $ \hat{Z}_c(k + j-1) \cap \mathcal{X}$ in CG-Rep.

As expected, the GI approach resulted in the highest set complexity since generators and constraints are added even if $ \hat{Z}_c(k + j-1) \subset \mathcal{X} $. The LP approach results in the lowest complexity by only adding generators and constraints when needed to exactly define the intersection. In this application, the ZH method also achieves this low set complexity and requires significantly less computation time. However, achieving this low complexity is not expected in general. Finally, the IA approach did not perform as well in this application, resulting in unnecessary generators and constraints and a large computation time. However, in practice, the zonotope-halfspace check from \textbf{Theorem \ref{zono_halfspace_int}} would be applied first so that \textbf{Algorithm \ref{gen_Bounds}} is only used in cases where the parent zonotope intersects the hyperplane. 

\IncMargin{1.5em}
\begin{algorithm2e}[t]
	\SetAlgoLined
	\SetKwInOut{Input}{Input}\SetKwInOut{Output}{Output}
	\Input{$\mathbf{x}^*$}
	\Output{$ Z_c(k)$}
	\BlankLine
	\SetAlgoLined
	initialize $ j \leftarrow N $\\
	$Z_c(k + j) =  \mathbf{x}^* $\;
	\While{$ j \geq 1 $}{
		$\hat{Z}_c(k + j-1) = A^{-1}Z_c(k + j) \oplus (-A^{-1}B) \mathcal{U}$\;
		$ Z_c(k + j-1) = \hat{Z}_c(k + j-1) \cap \mathcal{X}$\;
		$j \leftarrow j - 1 $\;
	}
	$Z_c(k) = Z_c(k + j) $
	\caption{Wayset $Z_c(k)$ for target $ \mathbf{x}^*$.}
	\label{waysetComputation}	
\end{algorithm2e}
\DecMargin{1.5em}

\begin{table} 
	\renewcommand{\arraystretch}{1.25}
	\caption{Complexity and Computation Time of Waysets}
	\label{Tab_wayset_cmplx_comptime}
	\centering
	\begin{tabular}{ M{1cm} | M{1.5cm} | M{0.8cm} | M{1.5cm} | M{0.8cm}}
		\multicolumn{5}{c}{} \\
		\hline	\multicolumn{1}{c}{} & \multicolumn{1}{c}{$Z_c$} & \multicolumn{1}{c}{$t_{calc}$} & \multicolumn{1}{c}{$\tilde{Z}_c$} & \multicolumn{1}{c}{$t_{calc}$} \\
		\cline{2-5}	\multicolumn{1}{c}{Method} & \multicolumn{1}{c}{$n_c \times n_g$} & \multicolumn{1}{c}{sec}
		& \multicolumn{1}{c}{$\tilde{n}_c \times \tilde{n}_g$} & \multicolumn{1}{c}{sec}\\
		\hline
		ZH & $ 7 \times 37$ & $1e^{-3}$ & 
		$7 \times 37$ & $4e^{-3}$ \\
		GI & $30 \times 60$ & $2e^{-3}$ & $7 \times 37$ & $2e^{-1}$ \\
		LP & $7 \times 37$ & $1e^{-1}$ & $7 \times 37$ & $2e^{-3}$ \\
		IA & $15 \times 45$ & $1e^{-1}$ & $7 \times 37$ & $4e^{-2}$\\
		H-Rep & $n_h = 5047$ & $161$ & $n_h = 153$ & $333$
	\end{tabular}
\end{table}

To demonstrate redundancy removal results from Section \ref{Sec_Redundancy}, \textbf{Algorithm \ref{waysetComputation}} and \textbf{Theorem \ref{Redundant_ConZono}} were applied to successfully remove all unnecessary generators and constraints resulting in the irredundant constrained zonotope wayset $ \tilde{Z}_c $ in Table \ref{Tab_wayset_cmplx_comptime}. Overall, when compared to H-Rep, any of the four CG-Rep approaches are computationally efficient with less set complexity and the preferred CG-Rep approach is likely to be application dependent.

\begin{figure}
	\begin{center}
 		\includegraphics[width=8cm]{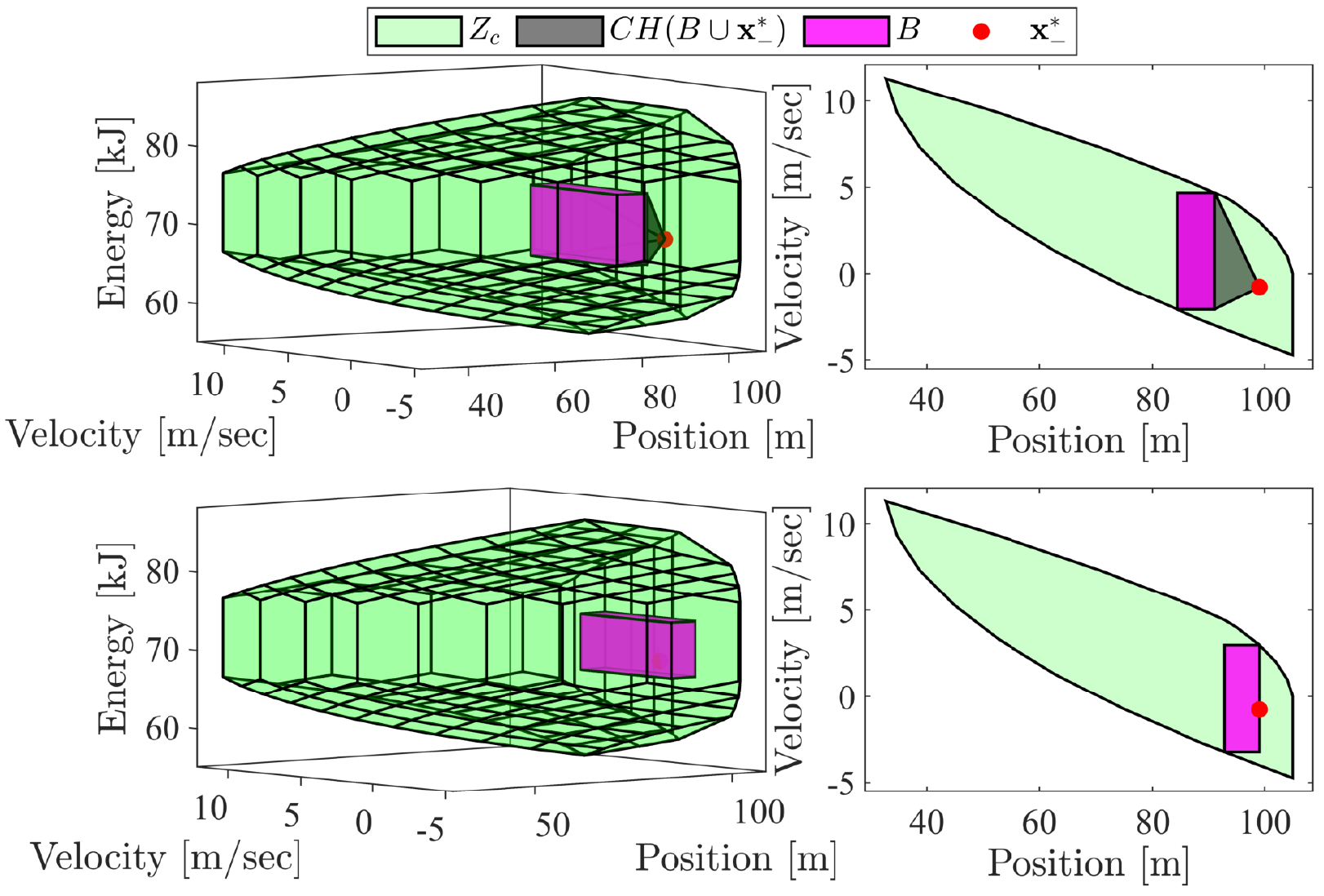}
		\caption{Top: The wayset $Z_c$, inner-approximating interval set $ B $ with $ V_r = 0.35 $, and $CH(B \cup \mathbf{x}^*_-)$ with $V_r = 0.39$ are shown on the left and the projections on to the position and velocity states are shown on the right. Bottom: The wayset $Z_c$, inner-approximating interval set $ B $ containing $ \mathbf{x}^*_- $ with $ V_r = 0.30 $ shown on the left with the projection shown on the right.}
		\label{Fig_HierMPC_InApprox_cvxhull}
		\end{center}      
\end{figure}

When computing these waysets for complex systems, it is likely that inner-approximations are needed to restrict the complexity of the set to satisfy a predetermined upper bound on the number of generators and constraints. Demonstrating the inner-approximations from Section \ref{Sec_InnerApprox} and the convex hull operation from Section \ref{Sec_ConvexHull}, the top row of plots in Fig. \ref{Fig_HierMPC_InApprox_cvxhull} shows the inner-approximating interval set $ B \subset Z_c $ computed using the method described in \textbf{Example \ref{example_innerApprox}} with $ n_g = 3 $ and $ n_c = 0 $.  However, in the hierarchical MPC framework from \cite{Koeln2019ACC,Koeln2019_Aut} the wayset must also include a key element denoted here as $ \mathbf{x}^*_- $. Since $ \mathbf{x}^*_- \notin B $, the wayset can be computed as $ CH(B \cup \mathbf{x}^*_-) $ resulting in $ n_g = 10 $ and $ n_c = 6 $. If this increase in set complexity is undesirable for a particular application, the point containment $ \mathbf{x}^*_- \in B \subseteq Z_c $ can be readily added to the LP defined in \eqref{Conzono_containment}. The resulting inner-approximating interval set with this point containment is shown in the bottom row of plots in Fig. \ref{Fig_HierMPC_InApprox_cvxhull}. The computation time for these inner-approximating interval sets are approximately $0.18$ and $0.25$ seconds for the top and bottom rows, respectively.

\section{Conclusions and Future Work} \label{Conclusions}
The use of zonotopes and constrained zonotopes for set operations provides significant computational advantages that improve the practicality of set-based techniques commonly used in systems and control theory. Operations such as halfspace intersections, convex hulls, invariant sets, and Pontryagin differences have been shown to benefit from zonotope and constrained zonotope set representations. 
Complexity reduction techniques were developed based on redundancy removal and inner-approximations to further improve the practicality of these set representations. Future work will focus on improved redundancy detection algorithms and optimization formulations that more accurately capture the volume of the approximated set.

\bibliography{library}
\balance

\end{document}